\documentclass[useAMS,usenatbib,usegraphicx]{mn2e}
\usepackage{comment}

\newif\ifAMStwofonts

\newcommand{\apjl}{ApJL}

\newcommand{\aap}{A\&A}

\newcommand{\br}{B$-$R }
\newcommand{\bv}{B$-$V }
\newcommand{\vr}{V$-$R }
\newcommand{\vi}{V$-$I }
\newcommand{\nodata}{-}

\begin{document}

\title[Globular clusters:  hosts and environment]
{The Connection between Globular Cluster Systems and their Host Galaxy and Environment:  A Case Study of the Isolated Elliptical NGC 821\thanks{Based upon data from the WIYN Observatory, which is a joint facility of the University of Wisconsin-Madison, Indiana University, Yale University, and the National Optical Astronomy Observatories.  Also includes analysis of observations made with the Hubble Space Telescope obtained from the ESO/ST-ECF Science Archive Facility.}
}
\author[L. Spitler et al.]
{
Lee R. Spitler$^{1}$\thanks{Email: lspitler@astro.swin.edu.au}, Duncan A. Forbes$^{1}$, Jay Strader$^{2}$,
Jean P. Brodie$^{2}$ and Jay S. Gallagher III$^{3}$\\ 
$^1$ Centre for Astrophysics \& Supercomputing, Swinburne University,
Hawthorn VIC 3122, Australia\\
$^2$ Lick Observatory, UC Santa Cruz, CA 95064, USA\\
$^3$ Department of Astronomy, University of Wisconsin-Madison, Madison, WI 53706, USA\\
}

\pagerange{\pageref{firstpage}--\pageref{lastpage}}
\def\LaTeX{L\kern-.36em\raise.3ex\hbox{a}\kern-.15em
    T\kern-.1667em\lower.7ex\hbox{E}\kern-.125emX}

\newtheorem{theorem}{Theorem}[section]

\label{firstpage}

\newpage

\maketitle

\begin{abstract}
In an effort to probe the globular cluster (GC) system of an isolated elliptical galaxy, a comprehensive analysis of the NGC~821 GC system was performed.  New imaging from the WIYN Mini-Mosaic imager, supplemented with HST WFPC2 images reveals a GC system similar to those found in counterpart ellipticals located in high density environments.  To put these results into the context of galaxy formation, a robustly-determined census of GC systems is presented and analysed for galaxies spanning a wide range of masses ($>${\it M$_{\ast}$}), morphologies and environments.

Results from this meta-study: (1) confirm previous findings that the number of GCs normalized by host galaxy {\it stellar mass} increases with host stellar mass.  Spiral galaxies in the sample show smaller relative GC numbers than those of massive ellipticals, suggesting the GC systems of massive ellipticals were not formed from major spiral-spiral mergers; (2) indicate that GC system numbers per unit galaxy {\it baryon mass} increases with host baryon mass and that {\it GC formation efficiency may not be universal} as previously thought; (3) suggest previously reported trends with environment may be incorrect due to sample bias or the use of galaxy stellar masses to normalize GC numbers.  Thus claims for environmentally dependent GC formation efficiencies should be revisited; (4) in combination with weak lensing halo mass estimates, suggest that GCs formed in direct proportion to the halo mass; (5) are consistent with theoretical predictions whereby the local epoch of re-ionization did not vary significantly with environment or host galaxy type.
\end{abstract}


\section{Introduction}

Observational evidence suggests galaxies in high-density regions of the Universe began to form stars at earlier epochs (redshifts of $z\sim5$ or $\sim11$ Gyr) and finish within shorter time periods (of roughly a few Gyr) compared to galaxies of low-density environments.  On average, galaxies in low-density environments started to form at $z\sim2$ and sustained low-level star formation rates even to the present day (Trager et al. 2000; Kuntschner et al. 2002; Proctor et al. 2004; Thomas et al. 2005).  What remains unclear are the exact conditions responsible for such an early divergence in the star formation histories of galaxies in high and low-density environments.  Globular clusters (GCs) are among the oldest ($>10$ Gyr) stellar systems in the Universe and their formation is associated with major star formation epochs (see the recent extragalactic GC system review of Brodie \& Strader 2006).  As such, systems of GCs provide unique observational tracers for investigating the environmental impact on galaxies at the onset of their formation (West et al. 2004).

Nearly all types of galaxies host GC systems that are bimodal in their observed optical colour distributions (see e.g. Peng et al. 2006; Strader et al. 2006).  Since GCs are mostly old, these bimodal colour distributions are interpreted as evidence for bimodal metallicity distributions, which may imply two major star formation epochs took place in {\it most } galaxies before the Universe was $\sim3$ Gyrs old.  Constraining the absolute age or the relative time between these two star formation epochs would provide invaluable constraints on galaxy formation, but is currently a formidable task (see e.g. Strader et al.~2005).

An alternative method for probing the early epochs is to determine the relative numbers of GCs across different galaxies.  This gives an indication of GC formation efficiency, which in turn helps characterise galaxy star formation histories as a function of galaxy mass, environment, and morphology.  The present work aims to characterise the GC system hosted by the nearby, isolated elliptical NGC~821.  Observational constraints gleaned from this GC system are an important contribution as nearly all previous work examined GC systems in the high-density environments of galaxy groups and clusters.

NGC 821 is an elliptical (E6) galaxy (M$_V = -21.0$; de Vaucouleurs et al. 1991) with a surface brightness fluctuation (SBF) distance modulus of $31.75\pm0.17$ ($\sim22$ Mpc; Tonry et al. 2001; corrected according to Jensen et al. 2003). It reveals a small edge-on disc both photometrically (Lauer 1985) and kinematically (Emsellem et al. 2004). Stellar ages range from 4.2 to 12.5 Gyrs in the literature, which Proctor et al. (2005) show to depend on the region of the galaxy sampled:  the central regions are $\sim$4 Gyrs old rising to 12.5 Gyrs at 1 effective radius. Proctor et al. concluded that the central starburst event represents only a few percent of the total galaxy mass.  Furthermore, they confirm a steep metallicity gradient and the alpha-element gradient was found to be flat with radius.  No cold gas component has been detected in NGC~821 (Georgakakis et al.~2001; Nakanishi et al.~2007).  Based on planetary nebulae kinematics, Romanowsky et al. (2003) claim NGC~821 has little, or no, dark matter out to $\sim4$ effective radii ({\it c.f.} Dekel et al. 2005 and Douglas et al. 2007).

After a detailed photometric analysis of the NGC~821 GC system, an in-depth comparison between this isolated GC system and those found in other environments is presented.  The implications for galaxy formation are discussed.

\section{Observations and Initial Data Reductions}

\begin{figure}
\resizebox{1\hsize}{!}{\includegraphics{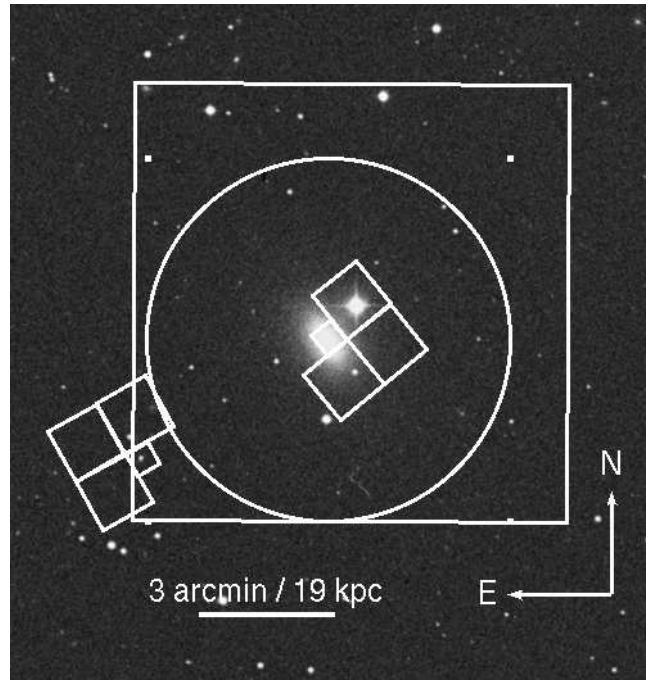}}
\caption{Digital Sky Survey image centred on NGC~821 showing the image footprints of the WIYN (square region) and two WFPC2 pointings (series of four boxes).  The circle illustrates the spatial extend of the GC system estimated in \S\ref{contamin}.}\label{figdss}
\end{figure}

Imaging of NGC~821 was acquired over three nights in late 2005 under good seeing conditions (average $~0.7"$) with University of Wisconsin-Madison time using the Mini-Mosaic ($9.6\arcmin\times9.6\arcmin$) imager installed on the 3.5-meter Wisconsin Indiana Yale NOAO (WIYN) Telescope.  The filters used (B, V, and R) and exposure times of the final image products are summarised in Table~\ref{tab1}.  For data reduction, the IRAF/PyRAF\footnote{IRAF is distributed by the National Optical Astronomy Observatories, which are operated by the Association of Universities for Research in Astronomy, Inc., under cooperative agreement with the National Science Foundation.  PyRAF is a product of the Space Telescope Science Institute, which is operated by AURA for NASA.} package MSCRED was used following the WIYN Mini-Mosaic online reduction notes by S. Kafka.  Figure~\ref{figdss} shows the orientation and coverage of the single WIYN pointing.  Observations were designed so that a 10th magnitude (B-band) star fell in the $7.12\arcsec$ gap between the two Mini-Mosaic CCDs, thus minimising severe saturation artifacts.  The native pixel scale of $0.141\arcsec$ pixel$^{-1}$ was maintained for the final science images.


\begin{table}
 \centering
  \caption{Exposure times for a given camera, target and filter.}\label{tab1}
  \begin{tabular}{@{}lr@{}}
\hline  
 WIYN & Obs. Date: 2005 Nov 30 - Dec 2  \\
\hline\hline 
NGC~821 - B &  9$\times$900s \\
NGC~821 - V &  4$\times$900s \\
NGC~821 - R &  5$\times$900s \\
\hline
 WFPC2 & HST Data: U307120*B \\
\hline \hline
NGC~821 - V & 4$\times$350s + 1$\times$120s\\
NGC~821 - I & 6$\times$230s + 1$\times$70s\\
\hline
 WFPC2 & HST Data: U4AN1[1-2]01B\\
\hline \hline
Background - V & 2$\times$2400s \\
Background - I & 2$\times$2400s \\
\hline
\end{tabular}
\end{table}

An archive HST WFPC2 dataset in V and I bands was also analysed to complement the WIYN images where the light from NGC~821 and the bright star significantly decreased object detection rates.  As shown in Figure~\ref{figdss}, this field is centred on NGC~821.  A second WFPC2 pointing that only just overlaps with the WIYN image is shown in the same figure.  This second dataset served as an estimate of the background contamination level for the central WFPC2 pointing.  Although the second pointing exhibits a fainter completeness limit than the WFPC2 pointing containing NGC~821, the only relevant analysis (characterising the GC luminosity function, see \S\ref{gclf}) robustly accounts for this difference.  The central WFPC2 image was downloaded from the ESO HST Science Association data archive.  The WFPC2 background pointing was produced manually with IRAF IMCOMBINE and designed to have a total exposure time similar to that of the central WFPC2 dataset.  Table~\ref{tab1} provides the effective exposure times of the WFPC2 datasets.

\section{Data Analysis}\label{dataanalysis}

\subsection{Outline}\label{outline}

The primary objective of this work is to investigate GC formation efficiency in an isolated elliptical.  More specifically, the number of NGC~821 GCs normalised by the mass of NGC~821 is a useful quantity to gauge GC formation efficiencies between different galaxies.  Accurately determining this value requires a robust estimate of the NGC~821 GC population, motivating the analytical procedure outlined here and carried out in the remainder of \S\ref{dataanalysis}.

Most GCs at the distance of NGC~821 will appear as point sources on everything but the small region covered by the high-resolution WFPC2-PC images and will therefore resemble Galactic stars and unresolved galaxies.  An objective selection criteria is therefore needed to remove these contaminates from the raw object dataset.  This proceedure will be described in \S$\ref{phot}$ and $\S\ref{gcselection}$.  Estimating the amount of contamination inevitably missed in this selection process is an important component to the GC number estimate (see $\S\ref{contamin}$).

Observational biases need to be quantified and considered during analysis.  For instance, the detection rates vary significantly with galactocentric distance due to the luminous centre of NGC~821 ($\S\ref{comptest}$).  Accounting for this motivates radial-dependant corrections to the GC luminosity function (GCLF; see $\S\ref{gclf}$) and the use of the central HST imaging.  To estimate the total number of metal-poor and metal-rich GCs, the varying proportion of these GC types again with galactocentric distance must be understood as well as any selection biases on colour ($\S\ref{colourdist}$).

This analysis comes together in $\S\ref{finalrad}$ and where robust population estimates of the NGC~821 GC system are presented.  Implications from this work are explored in \S\ref{results}.

\subsection{GC Photometry}\label{phot}

\subsubsection{WIYN}\label{wiynphot}
Sources on the WIYN images were detected on each of the three bands after a median filter (of $31\times31$ pixel window) image was subtracted from the original images, thereby removing the diffuse light of NGC~821.  Objects $3\sigma$ above the global background were catalogued with the {\it find} task in DAOPHOTII (Stetson 1992).  Flux from each object was extracted with DAOPHOTII {\it phot} in an aperture designed to keep the subsequent aperture corrections small.  Images still containing the diffuse NGC~821 light were used for the photometry.  Local background flux levels for individual objects was subtracted from the extracted flux and estimated from a 10-pixel wide annulus with an inner radius of 15 pixels (or $\sim2.1\arcsec$) from the object centre.  Aperture corrections were derived from point sources on the images.  Corrections for Galactic foreground (with a reddening excess of {\it E(B $-$ V)} $=0.11$; Schlegel et al. 1998) and airmass extinction were also incorporated into the total flux measurements.  Photometric zeropoints derived from standard Landolt (1992) field star observations were applied to the fluxes yielding the final magnitudes values on the standard Johnson-Cousin photometric system.  Photometric corrections are summarised in Table~\ref{tab2}.

\begin{table}
 \centering
  \caption{WIYN photometric corrections for the finite apertures, Galactic extinction and atmospheric reddening, each in units of magnitudes.  For each filter, the average airmass during observations was used for such corrections.}\label{tab2}
  \begin{tabular}{@{}lccccc@{}}
\hline
Filter & Aperture & Extinction & Airmass \\
  \hline \hline
B & $-0.14\pm0.04$ & $-0.48$ & $-0.22$ \\
V & $-0.13\pm0.04$ & $-0.37$ & $-0.14$ \\
R & $-0.19\pm0.03$ & $-0.30$ & $-0.08$ \\
\hline
\end{tabular}
\end{table}

\subsubsection{WFPC2}\label{wfpc2phot}

Both WFPC2 datasets (see Table~\ref{tab1}) were analysed following the aperture photometry procedure described in Larsen et al. (2001).  To this end, a generalised WFPC2 data reduction pipeline provided by S. Larsen was employed.  This pipeline makes use of the IRAF APPHOT package.  Objects deviating $3\sigma$ from the background noise level were found and photometry was measured in the central WFPC2 field from a median filter ($15\times15$ pixel window) subtracted image that effectively removed the diffuse light gradient from NGC~821.  For the background pointing, individual background values were computed for each object in annuli (aperture sizes as in Larsen et al. 2001) centred on each object.  All fluxes were calibrated to the standard photometric system according to the Holtzman et al. (1995) procedure and corrected for charge-transfer efficiency as presented in Stetson (1998).

Corrections for Galactic foreground extinction were applied with the same reddening excess value used for the WIYN photometry.  Aperture corrections to the $0\arcsec.5$ Holtzman et al. (1995) reference aperture were directly measured from the objects passing GC selection criteria (see $\S\ref{gcselection}$ for this criteria).  The median magnitude difference between the extraction and reference aperture of these candidate GCs on the WFPC2-WF CCDs was added to all fluxes to correct for light found beyond the extraction radius.  Separate aperture corrections for the WFPC2-PC CCDs were necessary because of its different point spread function.  There are not enough GC-like objects in the background WFPC2-PC CCD, thus a full extraction aperture of $0\arcsec.5$ was used for the 2 objects found within the WFPC2-PC field.  WFPC2 photometric corrections are summarised in Table~\ref{tab3}.

\begin{table}
 \centering
  \caption{WFPC2 photometric corrections for finite apertures and Galactic extinction.}\label{tab3}
  \begin{tabular}{@{}llcccc@{}}
\hline
Target & WFPC2 & Filter & Aperture & Extinct. \\
& Chip &  & Correction & \\
  \hline \hline
NGC~821 & PC  & V   & $-0.10$ & $-0.36$ \\
NGC~821 & WF  & V   & $-0.01$ & $-0.36$ \\
NGC~821 & PC  & I & $-0.17 $ & $-0.22$ \\
NGC~821 & WF  & I & $-0.19 $ & $-0.22$ \\
Background & WF  & V   & $-0.23$ & $-0.36$ \\
Background & WF  & I & $-0.25 $ & $-0.22$ \\
\hline
\end{tabular}
\end{table}

\subsection{Completeness Tests}\label{comptest}

\begin{figure}
\resizebox{1\hsize}{!}{\includegraphics{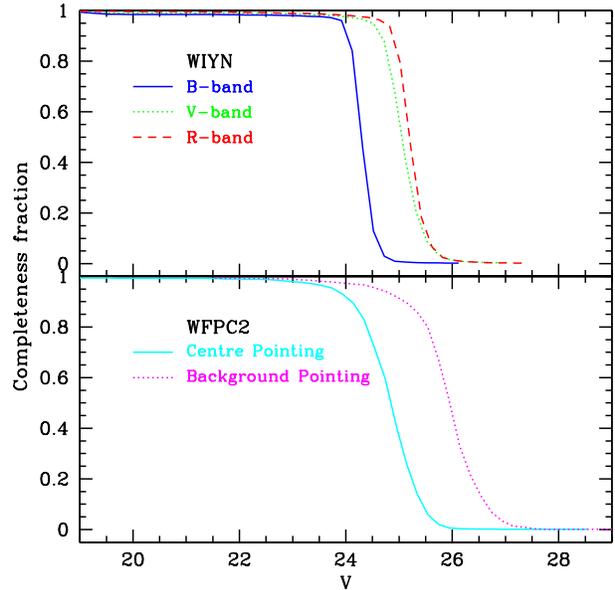}}
\caption{Completeness functions from tests described in the text.  {\it Top panel:} Results for the WIYN B-, V- and R-band images, represented by solid, dotted and dashed lines, respectively.  To facilitate this comparison, the B and R-band magnitudes were converted to V magnitudes with the mean GC colours of \bv $= 0.69$ and \vr $=0.45$.  Object recovery on the B-band image approaches the 50\% level for objects $\sim1$ mag. brighter than the V and R-bands.  {\it Bottom panel:}  WFPC2 completeness test results for the central and background pointings are represented by solid and dotted lines, respectively.}\label{figcomp}
\end{figure}

Completeness tests were performed to quantify the magnitude limits of each dataset.  Using the IRAF DAOPHOT package, artificial objects produced from empirical point spread functions were placed on to the original science images in such a way that crowding from existing and other artifical objects could not influence the test results.  For each 0.2 magnitude step between the expected magnitude range of GCs, 15000 and 500 objects of the same magnitude were added to the WIYN images and each WFPC2 CCD, respectively.  Since the WFPC2 data pipeline requires both a V and I-band image as input, the colour of the artificial objects was taken to be \vi $=1.1$.  This colour is typical of GCs (e.g. Kundu \& Whitmore 2001 and Larsen et al. 2001) The appropriate photometry extraction procedure (for WIYN see $\S\ref{wiynphot}$, WFPC2 $\S\ref{wfpc2phot}$) were carried out on each artificial field.  Completeness tests are given in Figure~\ref{figcomp}.   When appropriate, variations with galactocentric radius in the completeness function are considered in the subsequent analysis.

\subsection{Globular Cluster Selection Criteria}\label{gcselection}

\begin{table*}
 \centering
  \caption{Sample of the catalogue of astrometry and photometry for GC candidates detected around NGC~821 on the WIYN and central WFPC2 images.  The full table is available at the CDS.  Note those objects beyond a galactocentric radius of $4\arcmin$ from NGC~821 made up a contamination sample for the WIYN analysis (see $\S\ref{contamin}$).}\label{tabgccat}
  \begin{tabular}{@{}rlllllllllll@{}}
\hline
ID & RA (2000.0) & Dec (2000.0) & R$_{gc}$(arcmin) & $B$ & $\sigma B$ & $V$ & $\sigma V$ & $R$ & $\sigma R$ & $I$ & $\sigma I$ \\
  \hline \hline
   1 & 02:08:21.425 & 11:05:14.18 & 5.53 & 21.163 &  0.004 & 20.618 &  0.004 & 20.232 &  0.003 &      - &      -  \\
   2 & 02:08:20.657 & 10:58:55.60 & 0.79 & 21.741 &  0.012 & 20.764 &  0.006 & 20.312 &  0.010 & 19.618 &  0.005  \\
   3 & 02:08:25.373 & 10:59:53.71 & 1.06 & 21.806 &  0.007 & 21.096 &  0.007 & 20.633 &  0.006 &      - &      -  \\
\hline
\end{tabular}
\end{table*}

In the following two subsections, criteria are defined that effectively reduce the WIYN and WFPC2 raw object list to mostly GC candidates.  Part of the final GC catalogue is given in Table~\ref{tabgccat}, the full version of which is available from the authors or available online at the CDS.  For objects in common between the WIYN and WFPC2 datasets, the tri-band WIYN selection took precedence over the WFPC2 GC \vi selection, unless the object was $0.\arcmin4$ from NGC~821~in which case the WFPC2 \vi selection was used.  Aside from the data presented in Table~\ref{tabgccat} and the analysis where merging was necessary (e.g. determining GC subpopulation proportions, \S\ref{colourdist}, and the spatial distributions, \S\ref{n821gcs}), for the subsequent NGC~821 GC analysis, the WIYN and WFPC2 datasets were analysed independently of each other.

\subsubsection{WIYN}

\begin{figure}
\resizebox{1\hsize}{!}{\includegraphics{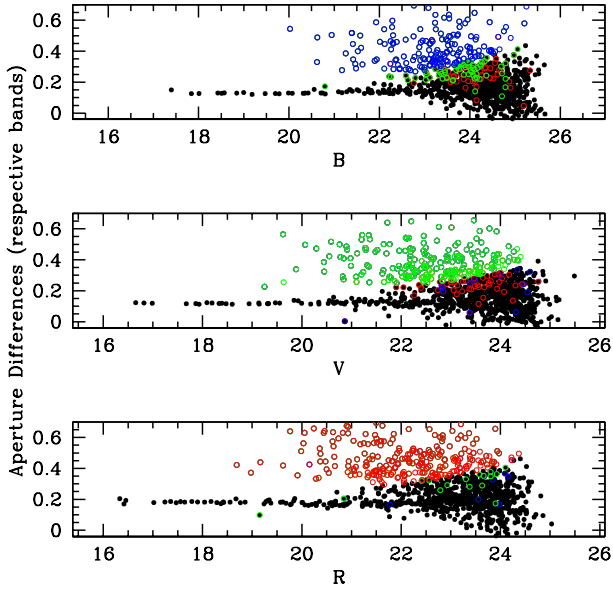}}
\caption{Selection criteria applied to remove extended sources (e.g. background galaxies) detected on the NGC~821 WIYN images.  The final B ({\it top panel}), V ({\it middle}), and R-band magnitudes ({\it bottom}) are plotted against the respective magnitude difference between the adopted extraction aperture and the one used to estimate the aperture corrections.  In each panel, those objects with the largest differences are considered extended sources and are represented as open circles, while point-source candidates are shown as solid circles.  In the online edition, the blue, green and red coloured open circles represent the flagged objects from extended object selection in the B, V and R-band photometry, respectively.}\label{figsize}
\end{figure}

Spurious detections such as image artifacts and statistical events were removed when individual objects were matched by their positions between the B, V and R-band WIYN images.  An extended-source culling of the remaining 1073 objects was carried out by removing objects that show an excess of light beyond the extraction aperture compared to point-sources (e.g. Holtzman et al.~1996).  This involves an analysis of the magnitude difference between the adopted extraction aperture and the outer aperture that defined the aperture corrections.  Figure~\ref{figsize} shows sequences of relatively bright point sources (most likely stars) that make up the fiducial point-source magnitude difference and define the aperture corrections.  Measurement uncertainties were accounted for by using a magnitude difference selection that depends on the final magnitude of the object.  The FWHM difference between point and extended sources suggests the size selection employed will remove any object at the distance of NGC~821 with a effective radius greater than $\sim9$~pc.  Less than 4\% of old Milky Way show sizes greater than 9~pc.  A total of 746 point-sources remain after the application of this aperture difference selection criteria.

\begin{figure}
\resizebox{1\hsize}{!}{\includegraphics{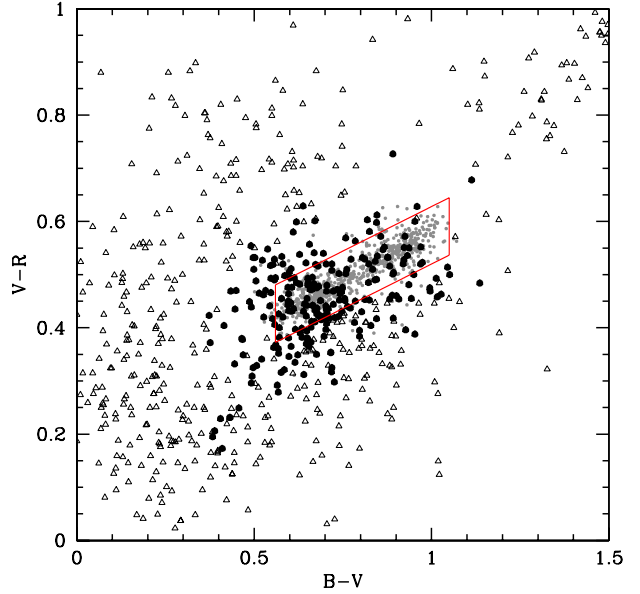}}
\caption{NGC~821 WIYN object \vr versus \bv plot.  Solid circles represent the final GC candidates, which fall within $2\sigma$ of the colour region defined by NGC~4594 GC HST photometry (small grey circles, Spitler et al. 2006).  This criteria was relaxed by an amount equal to an object's colour error, as described in the text.  Open triangles are WIYN objects failing these criteria and are probable stars or galaxies.}\label{figwiyncoloursel}
\end{figure}

A two-colour GC selection technique was used to further constrain the GC candidate list (e.g. Rhode \& Zepf 2001) to the same region occupied by known GCs in \bv versus \vr colour space.  Rhode \& Zepf (2001) used Galactic GCs as their basis sample.  Milky Way GCs have a relative deficiency of metal-rich GCs, thus the WIYN sample is instead compared to accurate HST photometry of 659 NGC~4594 GCs (Spitler et al. 2006), which are partly resolved in the HST images.  The NGC~821 and NGC~4594 datasets are presented in Figure~\ref{figwiyncoloursel}.  A linear relation was defined from the NGC~4594 GC colour-colour information and the final NGC~821 GC candidates are those with \vr colours falling within $2\sigma$ (where $\sigma$ is the intrinsic scatter of NGC~4594 GCs) of this relation.  \bv colour limits of 0.55 and 1.05 were also imposed and correspond to [Fe/H] values of $-2.5$ and $+0.2$ for Galactic GCs, according to a linear fit to the data in the Harris (1996) catalogue.  Following Rhode \& Zepf (2001), if an object's colour deviated from the criteria by an amount equal to its formal colour error, it was still considered a GC candidate.

Finally, objects were considered GC candidates if their V-band magnitude is fainter than V $= 20$ mags (i.e. M$_V=-11.75$).  This magnitude limit approximately corresponds to the boundary between GCs and more massive compact stellar objects (e.g. Hasegan et al. 2005; Mieske et al. 2006).  It is noted that three (plus one on the WFPC2 central pointing) point-sources with GC-like colours are brighter than V $=20$ mags.  These are candidate ultra-compact dwarf galaxies and are presented in Table~\ref{tabucd}.  Objects with photometric errors greater than 0.15 mags in any one band were also discarded, leaving 237 GC candidates from the WIYN images.

\begin{table}
 \centering
  \caption{Ultra-compact dwarf galaxy candidates.  The fourth entry was found on a WFPC2-WF chip, thus there is a chance the object is partially resolved.  Size measurements with IRAF/IMEXAM indicates no difference between the candidate and two bright stars on the same chip.  This puts a $\sim5$pc upper-limit on the effective radius of this UCD candidate.}\label{tabucd}
  \begin{tabular}{@{}cccccc@{}}
\hline
 R.A. & Dec. & V & \bv & \vi \\
  \hline \hline
02:08:05.135 &+10:57:35.13& $19.47$ & $0.96$ & \nodata \\
02:08:34.641&+11:00:38.03&$18.94$ & $0.54$ & \nodata \\
02:08:16.697&+11:01:24.51&$17.93$ & $1.07$ & \nodata \\
02:08:17.685&+11:00:31.84&$19.60$ & \nodata & $1.20$  \\
\hline
\end{tabular}
\end{table}

\subsubsection{WFPC2}

\begin{figure}
\resizebox{1\hsize}{!}{\includegraphics{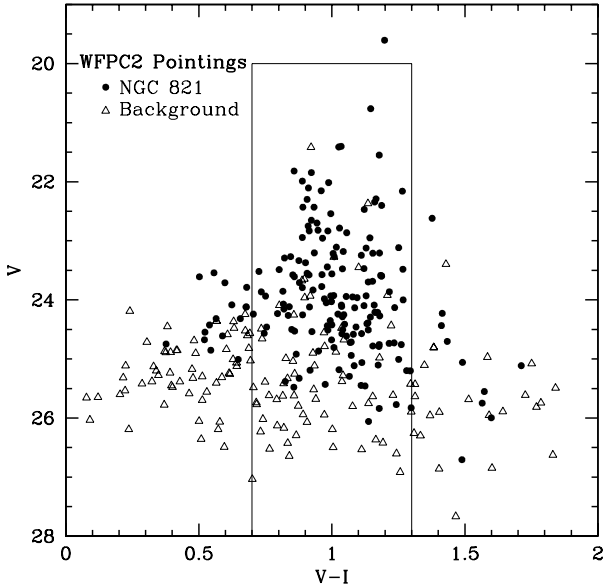}}
\caption{Colour-magnitude diagram of objects detected in the HST WFPC2 pointings.  Solid circles and open triangles represent point sources found in the central NGC~821 and background pointings, respectively.  GC selection limits are indicated with vertical lines at \vi $= 0.7 $ and $=1.3$ and a horizontal line at V $=20.0$.}\label{figwfpc2cmd}
\end{figure}

The adopted GC selection criteria for the two WFPC2 fields are identical.  Object with elongated or extended structure (determined by eye and with IRAF/IMEXAM) were culled.  Remaining objects detected on the WFPC2 fields are shown in the colour-magnitude diagram in Figure~\ref{figwfpc2cmd}.  GC colour selection of $0.7 <$ \vi $\le 1.3$ mags were applied and corresponds to roughly the same metallicity range as the WIYN \bv colour limits.  One object meeting these requirements was brighter than V $=20$ mags and was removed from further analysis, but is recognised as a candidate ultra-compact dwarf galaxy (see Table~\ref{tabucd}).  A total of 156 and 65 GC candidates remain after these criteria were applied in the central NGC~821 and background WFPC2 pointings, respectively.

\subsection{Contamination Estimates}\label{contamin}

\begin{figure}
\resizebox{1\hsize}{!}{\includegraphics{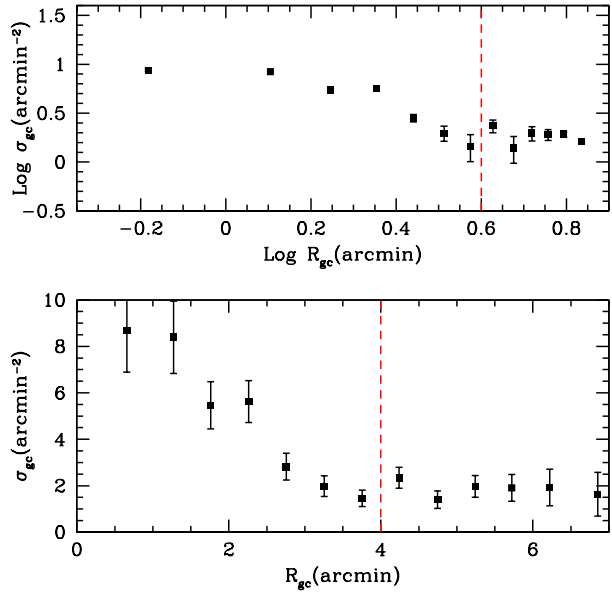}}
\caption{Raw WIYN NGC~821 GC surface density profiles.  The bottom panel shows a linear representation of the data and the top panel is in logarithmic form.  The consistency between the outer 6 points suggests the data reaches a constant background level by R$_{gc} \sim4$ arcmin ($=26$~kpc; indicated by the dashed vertical line).  Uncertainties are from Poisson statistics.}\label{figradinit}
\end{figure}

The amount of contamination in the WIYN NGC~821 dataset was estimated from the objects found in the region beyond the spatial extent of the GC system.  Figure~\ref{figradinit} shows the surface density profile from the raw counts of GC candidates in 12 radial annuli of equal widths ($0\arcmin.5$) centred on NGC~821.  Image boundaries in each band and masked regions where objects could not be detected (e.g. bright stars) were excluded in the annulus areal coverage calculations.  It is apparent in the figure that a constant background level of $1.5-2.0$ GC candidates per square arcmin begins at a projected distance of $\sim4$ arcmin ($4\arcmin=26$~kpc) from NGC~821.  This galactocentric radius (R$_{gc}$) is taken as the ``edge'' of the GC system, despite the probable existence of a few GCs beyond this nominal edge.  Figure~\ref{figdss} demonstrates that the GC system is effectively covered by the WIYN image, thus no further areal corrections are necessary for the GC population estimates.  The adopted GC system and contamination unmasked regions have total areal coverage of 48.6 and 41.4 square arcmin, respectively.

As shown in Figure~\ref{figdss}, the background WFPC2 pointing is also beyond the GC system edge, thus all objects passing the GC selection criteria in this pointing conveniently define the contamination level for the central NGC~821 WFPC2 pointing.  A combination of differing exposure times in these two datasets and the presence of NGC~821 in the central pointing undoubtedly leads to differing detection rates.  This is only relevant to characterising the GCLF (see \S\ref{gclf}) and so the GCLF analysis employed robustly accounts for this issue.

\subsection{Colour Distributions}\label{colourdist}

\begin{figure}
\resizebox{1\hsize}{!}{\includegraphics{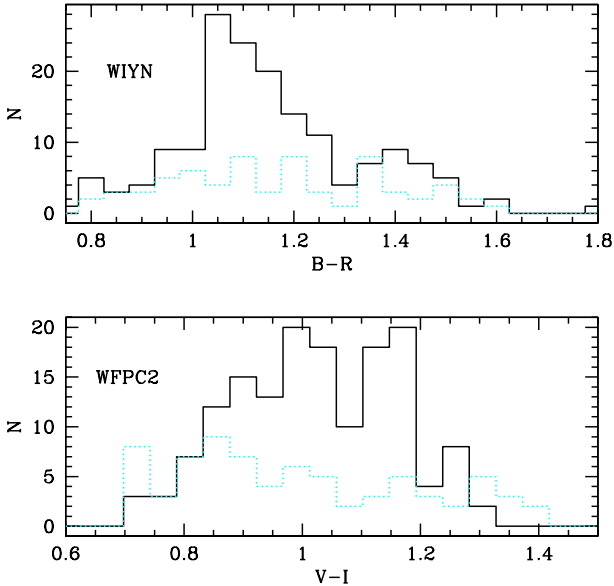}}
\caption{GC and background contamination colour histograms.  {\it Top panel:} Shows the \br histograms of candidate GCs (solid lines) and contamination objects (dotted lines) extracted from the WIYN pointing.  {\it Bottom panel: } \vi distribution from the central WFPC2 pointing and the contamination level estimated from the WFPC2 background image.  The split between the blue and red GCs occurs at \br $\sim1.3$ and \vi $\sim1.1$.  Blue GCs dominate the WIYN distribution, while the proportion of red GCs is nearly half the total in the central NGC~821 WFPC2 dataset.}\label{figcolourhists}
\end{figure}

Figure~\ref{figcolourhists} shows colour histograms of NGC~821 GC candidates in the WIYN and WFPC2 pointings.  GC colour bimodality is most apparent in the central WFPC2 pointing, while the WIYN colour distribution shows only a slight increase over the estimated contamination level at colours consistent with metal-rich GCs.

A KMM test for bimodality (Ashman et al.~1994) suggests a marginal probability of bimodality in the WFPC2 distribution (with a p-statistic of 0.0651) and no statistically significant evidence (p-statistic of 0.6672) for bimodality in the WIYN \br distribution.  Two important facts must be considered before interpreting these KMM results:  the established observational property of GC systems that red (metal-rich) GCs tend to be more centrally concentrated (e.g. Geisler et al.~1996) and a poor detection rate near NGC~821 in the WIYN images.  The first point leads to a detection bias towards red GCs on the WFPC2 image, while the combination of the two biases the WIYN detections against red GCs.  Overcoming these biases requires combining the WIYN and WFPC2 GC colour information, thereby using the WFPC2 data to ``fill-in'' the central region where detections fall on the WIYN image, near NGC~821.  The WIYN \br GC colours were therefore converted to the WFPC2 \vi colours with an empirical conversion from GC candidates in common between the datasets:  

\begin{equation} \vi=0.24\pm0.14+0.65\pm0.11(\br) \end{equation}

Objects in common between the WIYN and WFPC2 datasets were only counted once.

Two additional observational biases must also be accounted for when robustly characterising the GC colour distribution.  First, the combined colour data were corrected for colour biases arising because a single band limits the sample, which spans a range of colours.  The following faint limits therefore need to be enforced so that GCs of all colours were detected at the $>95\%$ completeness level:  WIYN R $<23.0$ mag and WFPC2 V $<23.5$.  Such a restriction also reduces uncertainties associated with contamination, which is generally faint.  Second, metal-rich GCs have higher mass-to-light ratios due to emission line blanketing, so these GCs should be fainter (by B $=0.3$ and R $=0.2$ mags; Ashman et al. 1995) compared to a metal-poor GC of identical mass.  Colour-dependent magnitude selections were therefore applied:  V$_{WFPC2}<3/2$ \vi $+21.8$ and R$_{WIYN}<2/3$ \br $+22.4$.

\begin{figure}
\resizebox{1\hsize}{!}{\includegraphics{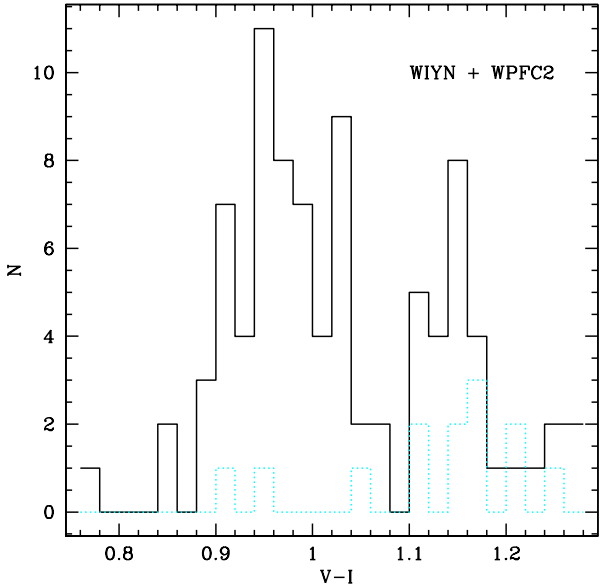}}
\caption{Histogram of combined WIYN and WFPC2 \vi colour values.  Solid and dotted histograms are for the GC candidates and objects in the background contamination fields, respectively.  See text for magnitude limits imposed to remove colour biases.  WIYN \br photometry was converted to WFPC2 \vi with an empirical relation derived from common objects.  Output from a KMM analysis of this data strongly supports the apparent bimodality and indicates $\sim30\%$ of the GCs in NGC~821 can be associated with the red subpopulation.  A Monte Carlo simulation supports GC colour bimodality at the $1\sigma$ level.}\label{fighistvi}
\end{figure}

The combined GC data, corrected for these biases (totalling 87 objects), is presented in Figure~\ref{fighistvi}.  A bimodal colour distribution is apparent:  a large, ``blue'', peak in the histogram is clearly observed at \vi $\sim1$ and a smaller ``red'' peak is centred at \vi $\sim1.15$.  Results from KMM analysis on this combined colour dataset suggest unimodality can be rejected at a confidence level of 99.998\% (with a p-statistic of 0.00160), strongly supporting the apparent \vi bimodality in the figure.  The KMM estimate for the two colour peaks is \vi $=0.99$ and $=1.18$, where the colour peak uncertainties are likely dominated by the zeropoint errors associated with the conversion from B--R to V--I colours:  0.14 mag.  

According to the KMM test, approximately 30\% of the total falls into the red GC subpopulation.  Figure~\ref{fighistvi} also shows the combined background contamination datasets, which similarly restricted in magnitude-space to account for the two biases detailed above.  The red (\vi $>1.1$) contamination level corresponds to approximately half the number of red GCs in the Figure, while the fraction of estimated contamination among the blue GCs is significantly lower.  This perhaps indicates the proportion of red GCs is somewhat smaller than the 30\% derived with the KMM test, suggesting this value is actually an upper limit.

To test the robustness of these KMM results as they relate to the uncertainty associated with the \br to \vi colour conversion (see Eq.~1), a 500-run Monte Carlo simulation was performed.  The WIYN objects were converted to \vi colours using 500 different linear relations derived from randomly deviating the zeropoint and slope of Equation~1 by their formal errors.  According to a KMM analysis of the simulated datasets, a strong rejection of unimodality bimodality (i.e. a KMM p-statistic $<0.05$; Ashman et al. 1994) is found in $\sim80\%$ of the runs, marginal evidence for bimodality occurs in 5\% and the remaining 15\% show no bimodality.  Thus, a strong case for colour bimodality in the NGC~821 GC system is supported at a $1\sigma$ confidence level.

\subsection{Globular Cluster Luminosity Function}\label{gclf}

The mean magnitude of a particular GC system is fairly constant for all galaxies for which it has been observed (e.g. Harris 2001; Richtler 2003; Jord{\'a}n et al. 2007b).  The shape of the GCLF can be approximated with a Gaussian distribution and in principle, the Gaussian ``peak'' or turnover magnitude (TOM) of a GCLF provides a useful quantity for comparison between GC systems.  Richtler (2003) aggregated WFPC2 GCLF TOM values and produced a ``universal'' TOM of M$_V=-7.35$ (here adjusted for the Jensen et al.~2003 SBF correction) with an observed RMS of $0.25$ mag.  Observational evidence suggests the GCLF does not vary with R$_{gc}$ (see refs. in de~Grijis \& Parmentier 2007), thus the GCLF in the central WFPC2 region should be similar in form to the GCLF derived from the surrounding WIYN coverage.

Secker \& Harris (1993) designed a maximum-likelihood technique to characterise the GCLF.  This program corrects for magnitude incompleteness, contamination, and compensates for a different completeness function for the contamination dataset.  It produces a best-fitting GCLF TOM and dispersion assuming either a Gaussian or Student t$_5$ ($t_5$) distribution.  Completeness functions and contamination levels differ between the WIYN and WFPC2 datasets, thus the analysis in this section makes use of the separate GC candidate lists from the respective images.

\begin{figure}
\resizebox{1\hsize}{!}{\includegraphics{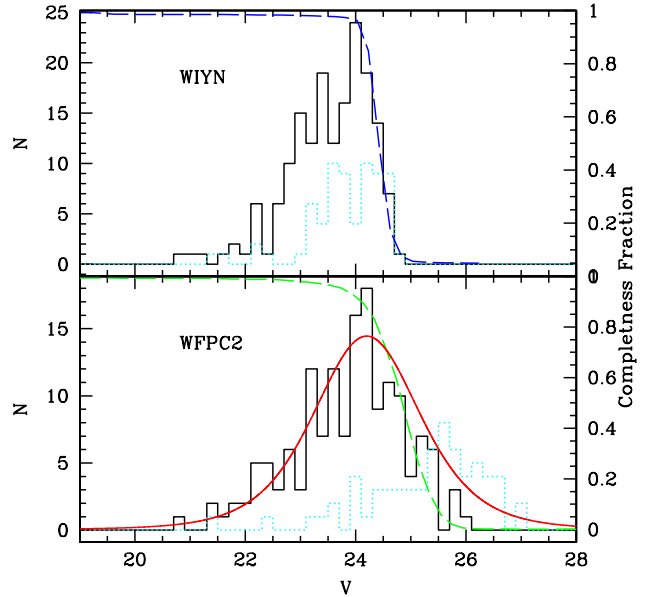}}
\caption{V-band NGC~821 GC candidate magnitude distribution.  For each panel, the left axis is for the histograms and the right is for the completeness functions (dashed lines).  WIYN and WFPC2 datasets are presented in the top and bottom panels, respectively.  Solid histograms are GC candidates and dotted histograms are the contamination datasets.  The best-fit $t_5$ distribution is shown over the WFPC2 histogram.}\label{figgclf}
\end{figure}

Assuming the Richtler (2003) universal TOM applies to NGC~821, GCs at the TOM magnitude should show an apparent magnitude of V$\sim24.4\pm0.3$.  Figure~\ref{figgclf} shows that the completeness level at this magnitude is 45\% and 75\% for the WIYN, and WFPC2 datasets, respectively.  Because only a fraction of the NGC~821 GCs at the expected TOM are actually observed, the uncertainty associated with the GCLF parameters from the Secker \& Harris program increases by a non-negligible amount, as described in Secker \& Harris (1993).  Nevertheless, applying this program to the WFPC2 GC candidate and contamination datasets, assuming a $t_5$ form of the GCLF (which provides a better fit than a Gaussian), yields the following parameters:  V$_{TOM}=24.2\pm0.3$ and a dispersion of $\sigma_{t5}=1.0\pm0.2$.  A Gaussian fit produced the following values: V$_{TOM}=24.3\pm0.4$ and $\sigma=1.2\pm0.2$.  The best-fit TOM values are consistent with the universal TOM.

The Secker \& Harris (1993) program also estimates the number of GCs in the corrected, best-fit distribution.  The WFPC2 $t_5$ fit corresponds to N$_{GC}=130\pm20$ {\it within} the WFPC2 field of view.  Estimating the {\it total} number of GCs in NGC~821 requires the extension of this analysis to larger galactocentric radii and hence the use of WIYN GC candidates.  Unfortunately, because the WIYN 50\% completeness limit approximately corresponds to the estimated GCLF TOM, the solution of the Secker \& Harris algorithm is much too uncertain for any practical use.  Thus, a modified version of the Secker \& Harris (1993) program was designed to produce a GC number estimate for a specified GCLF TOM and dispersion.  GCLF parameters derived from the WFPC2 data were assumed and applied to the WIYN data as described in the following paragraph.

The B-band image limits the WIYN photometry (see $\S\ref{comptest}$), thus the input completeness function to the Secker \& Harris program was the B-band completeness function converted to V with the mean GC colour: \bv $=0.69$.  Since the number of GCs and the WIYN completeness function both vary significantly with R$_{gc}$, the WIYN GCLF fits were performed on the data in $0\arcmin.5$-wide annuli.  Appropriate completeness functions are used for each R$_{gc}$ interval and the final GC surface density estimates are given in Table~\ref{tab4}.  Similar analysis of the WFPC2 data yielded the two inner-most surface density values.  Surface densities errors are from Poisson statistics added in quadrature with the uncertainty in GC number estimate.  The latter value is dominated by the measurement uncertainty of the TOM, which is easily translated to a GC number error with the modified Secker \& Harris program.

\begin{table}
 \centering
  \caption{Robust GC number density estimates from GCLF fits and the relevant masked image areas.  Radii are the projected mean galactocentric distance from NGC~821 for a given annulus.}\label{tab4}
  \begin{tabular}{@{}lrr@{}}
\hline
Radius & $\sigma_{GC}$ & Unmasked Area \\
 (arcmin) & (arcmin$^{-2}$) & (arcmin$^{2}$) \\
  \hline \hline
0.2&$144.\pm38.$&0.32\\
1.1&$20.4\pm10.$&4.74\\
1.3&$11.9\pm3.1$&3.45\\
1.8&$6.40\pm1.8$&5.31\\
2.3&$6.78\pm1.8$&6.94\\
2.8&$2.00\pm0.8$&8.51\\
3.3&$0.50\pm0.5$&10.1\\
3.8&$0.00\pm0.4$&11.7\\
\hline
\end{tabular}
\end{table}

\subsection{GC Radial Profile and Population Estimates}\label{finalrad}

\begin{figure}
\resizebox{1\hsize}{!}{\includegraphics{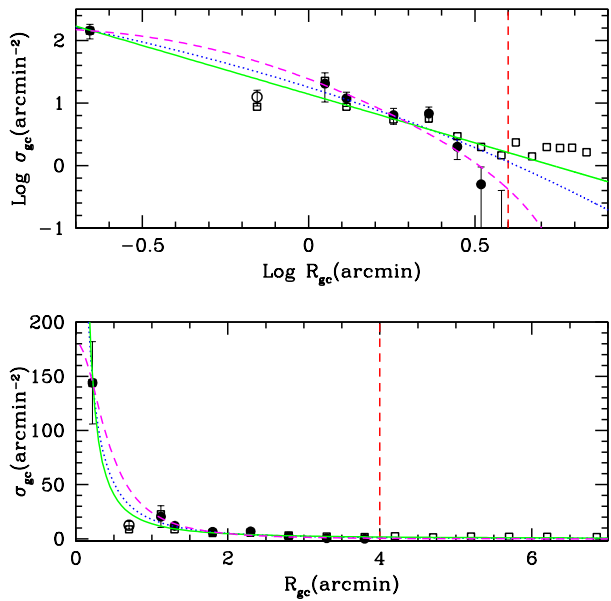}}
\caption{GC surface density corrected for contamination and incompleteness.  Uncorrected data points are given as open squares.  Circles are the corrected surface density from WIYN data, except the two inner-most solid points are from the WFPC2 data.  The open circle is not fit due to the high incompleteness fraction near NGC~821 on the WIYN image.  The outer six points define the contamination level.  Formal errors are from Poisson statistics and GCLF measurement uncertainties added in quadrature.  Best-fit power-law, de~Vaucouleurs and King (1962) functions are given as solid, dotted and dashed lines, respectively.}\label{figcorsurfdens}
\end{figure}

The corrected NGC~821 GC surface density profile from data values in Table~\ref{tab4}, is presented in Figure~\ref{figcorsurfdens}.  To derive an estimate of the GC population total that accounts for completeness fraction variations with galactocentric radius, a function is fit to the surface density data, which is then integrated with respect to the area covered by the NGC~821 GC system.  Three functions are fitted:  a power-law ($\sigma_{GC}\propto$R$^\alpha$), de~Vaucouleurs profile ($\sigma_{GC}=10^{\alpha_{deV} R^{1/4} +\beta_{deV}}$) and King (1962) profile ($\sigma_{GC}=\alpha_{king}[1/(1+r^2/\beta_{king}^2)^{1/2}-1/(1+15^2)^{1/2}]^2$).  A glance at the fits, presented in Figure~\ref{figcorsurfdens}, gives the impression that the King profile best represents the GC number density profile.  However, both the de~Vaucouleur and King fits yield consistent GC population estimates when surface integrated to the GC system's edge:  330 and 335, respectively.  The former profile is used in the subsequent analysis since results derived from it are directly comparable to past studies (e.g. Rhode \& Zepf 2004).

To test the significance of the de~Vaucouleur number estimate, a Monte Carlo simulation was performed where the observed surface density points were randomly deviated by their formal errors (assuming Gaussian errors).  The resulting values were fitted, and the new profile was integrated as before.  Repeating this 5000 times resulted in a distribution of population numbers.  The mean of the distribution is taken to be the final GC population total:  N$=320\pm45$, where the quoted uncertainty is the standard deviation of the simulated number distribution.  The present NGC~821 GC system {\it total} is smaller than the Kundu \& Whitmore (2000) number (N$=395\pm94$) they estimated {\it within} the central WFPC2 field of view.

Analysis of the GC colour distribution ($\S\ref{colourdist}$) suggests 30\% of the GCs are red, thus the NGC~821 GC system contains approximately $95\pm15$ red and $225\pm30$ blue GCs.

\section{Results and Discussion}\label{results}\label{conclusions}

\subsection{The GC System of NGC~821}\label{n821gcs}

GC specific frequency (S$_N$) has been used for comparing the abundance of GCs across different galaxies (Harris \& van den Bergh 1981) and is defined in Table~\ref{variables}.  The NGC~821 GC system population estimated in $\S\ref{finalrad}$ (N$_{GC}=320\pm45$) corresponds to a V-band specific frequency of S$_N = 1.32\pm0.15$.  This value is slightly lower compared to values traditionally associated with low-luminosity elliptical galaxies (e.g. Harris 2001).  However, recent efforts have yielded generally smaller S$_N$ values when a careful accounting for contamination is performed (e.g. Rhode \& Zepf 2004).  Indeed, it will be shown in $\S\ref{tbluemass}$ that the NGC~821 stellar mass-normalised number of GC associated is roughly consistent with the same values for similar-mass elliptical and spiral galaxies.

The NGC~821 GC colour distribution shows a hint of the classical bimodal form, with a fairly typical fraction ($30\%$) of red GCs.  The subpopulation mean colour values (\vi $= 0.97$ and $1.18$ mags.; errors $\sim0.14$) are roughly consistent with the same values in galaxies of similar luminosities (Larsen et al. 2001; Strader, Brodie \& Forbes 2004).  While a young ($\sim1-4$ Gyr) central stellar population in NGC~821 was confirmed by Proctor et al. (2005), they demonstrated this component only reflects $\le10$ per cent of the mass in the central regions of NGC~821.  Beyond one effective radius, Proctor et al. (2005) found consistently old ages ($\sim12$ Gyr) from the integrated spectrum of NGC~821.  With no evidence for a major star formation event in 12 Gyr, it is likely the majority of NGC~821 GCs are at least as old, as is believed to be the case for most GC systems (see references in Brodie \& Strader 2006).  The GC colour distribution can therefore directly be interpreted as evidence for a bimodal metallicity distribution, here shown with marginal certainty in NGC~821 for the first time (see Fig.~\ref{fighistvi}).

\begin{figure*}
\resizebox{1\hsize}{!}{\includegraphics{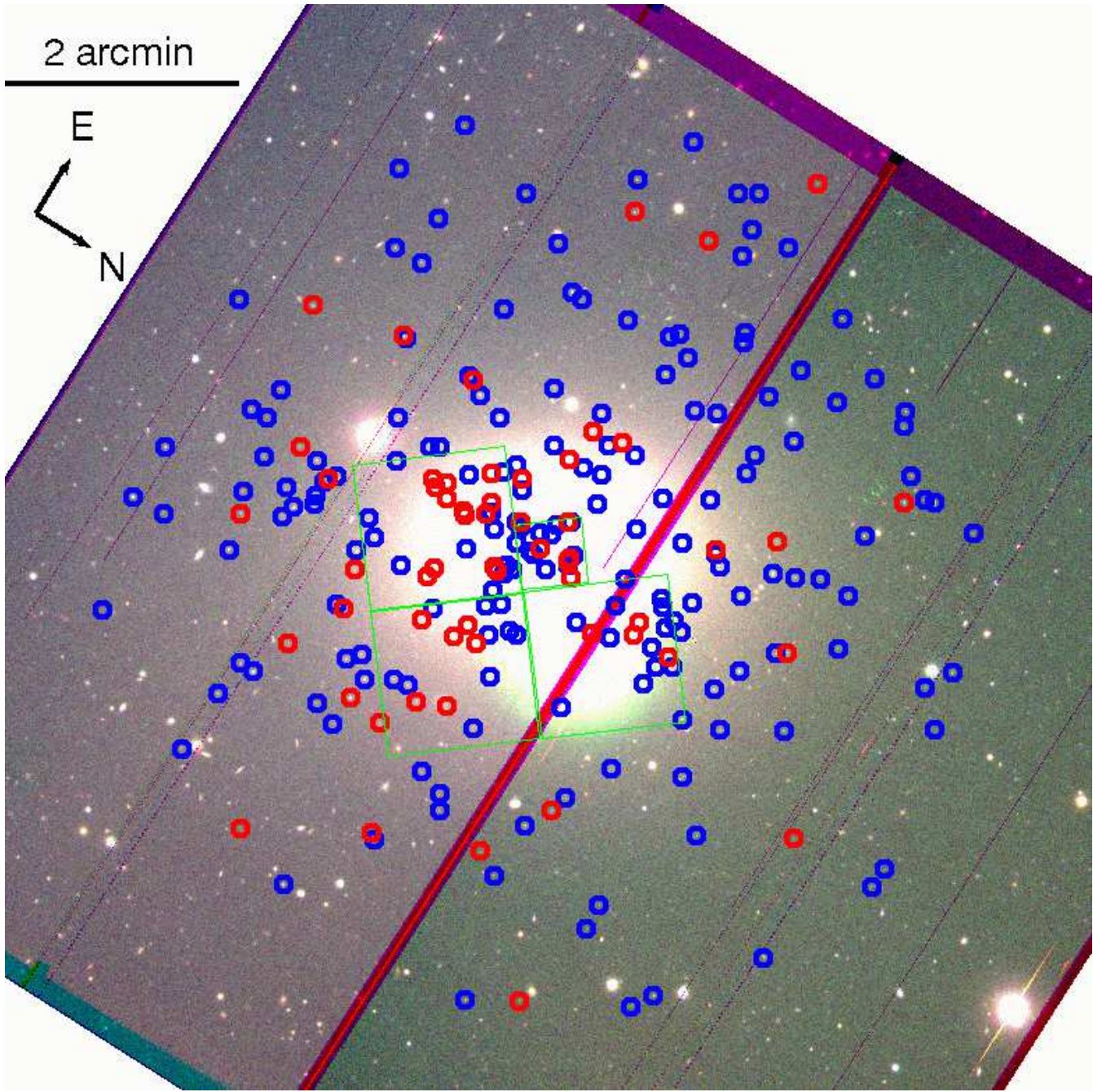}}
\caption{WIYN colour composite image showing the locations of GC candidates (blue and red circles).  The central WFPC2 image coverage is given.  The image is oriented so that the major-axis of NGC~821 is parallel with the width of the page.  A deficiency of GCs is apparent along the galaxy's minor-axis, which does not result from selection effects due to the bright foreground star immediately below the figure centre.}\label{figwiyn}
\end{figure*}

Red GCs in Figure~\ref{figwiyn} tend to be more centrally concentrated, as is commonly found in GC systems.  The GC system extends to a projected galactocentric radius of R$_{gc} \sim4$ arcmin or $\sim26$~kpc, which is a factor of $\sim5$ greater than the effective radius of NGC~821 ($50\arcsec$; de Vaucouleurs et al. 1991).

In the region north-west of NGC~821 (see Figure~\ref{figwiyn}; along the galaxy minor-axis) there is an apparent under-density of GC candidates relative to other azimuthal directions.  While the 10th magnitude foreground star (also just north-west of NGC~821) significantly influences detection rates on the WIYN images, the near-normal detection rates on the WFPC2 images sufficiently compensated for this ruined WIYN region.  Thus, with no significant azimuthal bias, the low numbers of GCs perpendicular to the major-axis of NGC~821 must be treated as a real trend, although small number effects are not yet ruled out.

\begin{figure}
\resizebox{1\hsize}{!}{\includegraphics{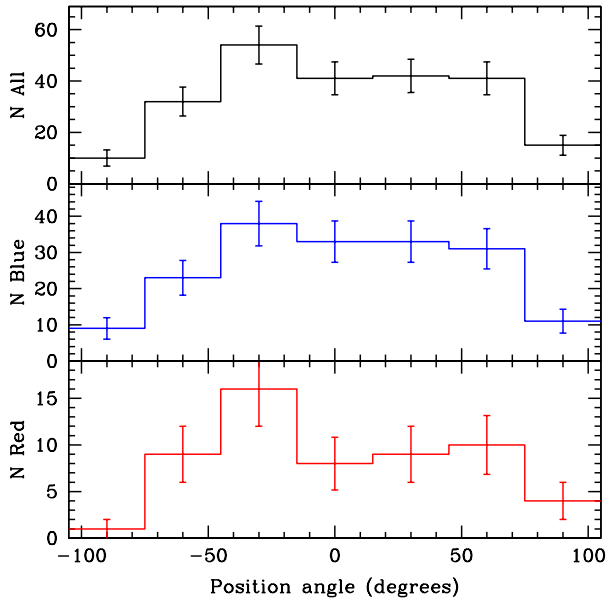}}
\caption{Folded position angle histograms of individual GCs associated with the whole system ({\it top panel}), blue ({\it middle}), and red ({\it bottom}) subpopulations.  Errors are Poisson statistics.  A PA of $0^{\circ}$ corresponds to the direction along the galaxy's major-axis.  A deficiency of GCs along the minor-axis is observed.}\label{figpa}
\end{figure}

Folded position angles (PAs) of individual GCs in the whole system and two subpopulations are presented in Figure~\ref{figpa}.  A PA of $0^{\circ}$ and $90^{\circ}$ (and $-90^{\circ}$) corresponds to directions along the galaxy major and minor-axis, respectively.  Data in Figure~\ref{figpa} confirms the spatial distribution of GCs in Figure~\ref{figwiyn} and apparent like orientation between the GC system and major-axis of NGC~821.  This is consistent with previous observations (see summary in Brodie \& Strader 2006) where GC systems generally follow the shape of the spheroidal component of their host galaxy.  For NGC~821 the situation is likely more complex, as the dearth of GCs is most apparent in the direction north-west of NGC~821, and not so obvious to the south-east.  It should be noted, there is also a hint of a major-axis orientation among the planetary nebula analysed by Romanowsky et al. (2003; see their fig. S2).  GC kinematic information would provide a valuable extra dimension to this analysis.

Information gleaned from the NGC~821 GC system is generally consistent with a scenario where this galaxy passively evolved since the formation of the bulk of the field stars (Proctor et al. 2005) and GCs roughly $12$ Gyr ago.  More specifically:  a relatively steep radial profile is inconsistent with a dissipation-less major merger (Bekki \& Forbes 2006) and evidence for a standard bimodal GC colour distribution is consistent with no late ($<5$ Gyr) major GC formation epochs.  Mild nuclear activity (Pellegrini et al. 2007), negligible cold gas components (from limits on HI content and no CO emission detections:  Georgakakis et al.~2001; Nakanishi et al.~2007) further support the idea that only small amounts of star formation took place during the recent past of NGC~821 (Proctor et al. 2005).  An extended, but relatively subdued formation history is consistent with the conclusions from a spatially-resolved spectroscopic analysis of 12 isolated early-type galaxies by Reda et al. (2007).

\subsection{A New Compilation of Globular Cluster Systems}\label{datacat}

To probe the relationship between GC system formation and environment a literature search for reliable GC system properties was conducted.  The compilations of Rhode, Zepf \& Santos (2005) (hereafter referred to as RZS05) and Brodie \& Strader (2006) served as the primary, but not exclusive source of literature references.  Inclusion criteria are the same as in RZS05.  Included GC system studies have:  data for 50\% of the GC system radial extent, observed in at least two filters, total GC number and subpopulation proportion given or derivable, and a GC number estimate error less than 40\%.  Exceptions to the radial coverage restriction were given to NGC~5128 and M87, as these pose difficult observational targets, but were analysed sufficiently well (see Harris et al. 2006 and Tamura et al. 2005ab) to be included in the present compilation.  Table~\ref{datatable1} contains this new compilation and Appendix~\ref{appendix} contains detail of its construction and references.  It also provides a procedure to estimate stellar masses from the absolute K or V-band total magnitudes of a galaxy, given its Hubble Type.

The local environmental galaxy density parameter ($\rho$; units of Mpc$^{-3}$) from Tully (1988a) is included in Table~\ref{datatable1}.  A smaller smoothing scale ($\sigma=0.5$) than employed in the original catalogue is used to increase the density resolution.  Density $\rho$ values ranging from 0--1, 1--5, 5--10, and $>10$ Mpc$^{-3}$ generally correspond to field, small group, large group (or small cluster), and cluster environments, respectively.

Errors on the $\rho$ values were approximated with the following formula:  $\Delta\rho=|(F \rho)-\rho|/\rho$, where $F($m--M$)$ is from Tully~(1988b) and is the expected number of galaxies over the number actually observable due to an increasing incompleteness fraction with distance.  The average uncertainty of the present sample is $\Delta\rho\sim0.7$.  The Tully (1988a) catalogue relies on recession velocities for distances, therefore individual peculiar motions could influence $\rho$, especially in low-density environments where the low counts are prone to small number statistics.

NGC~821 one of the most distant galaxies in the present compilation, thus the corresponding uncertainty in $\rho$ ($\rho=0.32\pm2.0$) is too large to prove its isolated nature.  However, confirmation of the isolated environ of NGC~821 was independently determined by Reda et al. (2004; and others prior, referenced in Reda et al.) from the larger LEDA galaxy catalogue (Paturel et al. 2003).  Since the Tully~(1988a) NGC~821 $\rho$ is likely under-estimated due to catalogue incompleteness, an upper-limit of $\rho=0.95\pm0.5$ was assumed.  This value is the Tully~(1988a) density value of the nearby galaxy NGC~4594, which, according to an appropriate NED query, meets the isolated criteria of Reda et al. (2004).  Other galaxies with distance moduli greater than m--M$=31.5$ (NGC~1407, NGC~3268 and NGC~3258) were omitted from the density analysis as the $\rho$ values are very likely to be underestimated.

\begin{figure}
\resizebox{1\hsize}{!}{\includegraphics{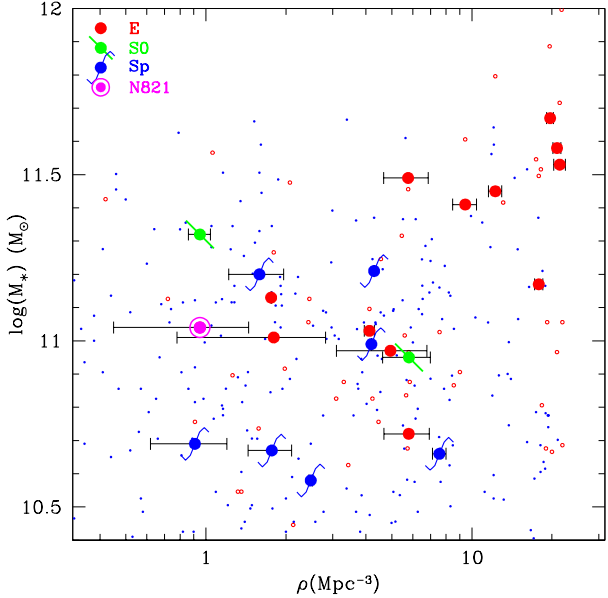}}
\caption{Galaxy stellar mass versus local galaxy density from Tully (1988a).  Large symbols representing galaxies with GC systems studied are as indicated in the figure legend.  Small blue points and open red circles are nearby (m--M$<31.5$) Tully (1988a) catalogue spiral and elliptical galaxies, respectively.  A comparison between the two datasets suggests observational biases are evident in the present sample:  GC systems of massive galaxies are preferentially studied in high-density environments.  Galaxy stellar masses are from K-band photometry (see Appendix~A for details).  Errors on $\rho$ reflect incompleteness fractions in the Tully (1988a) catalogue (see text).}\label{figbias}
\end{figure}

Many GC system properties are known to correlate with the mass of the host galaxy.  It is therefore important to understand selection biases relating to galaxy mass and environment.  Indeed, Figure~\ref{figbias} shows that the current sample is strongly biased towards very massive galaxies in the highest density environments.  Part of this results from the preferential examination of the most populous GC systems in order to compensate for the relatively large distances to galaxy clusters, but also reflects the fact that high-density regions are more likely to contain the most luminous galaxies.

Before presenting the results, the use of the GC T-parameters (Zepf \& Ashman 1993) instead of the galaxy luminosity-normalised GC specific frequency (S$_{N}$) should be motivated.  The T-parameter quantity, T$_N$, is defined as the number of GCs normalised by its host galaxy stellar mass (see Tab.~\ref{variables}).  The T$_N$ carries more physical meaning than S$_N$ because it is designed to account for stellar population differences (e.g. an old elliptical versus a young spiral galaxy) that bias optical photometry traditionally used for S$_N$ calculations (Zepf \& Ashman 1993).

To improve clarity in the following discussion, two new quantities are introduced as galaxy baryon and halo mass analogues to the S$_N$ and T$_N$ quantities.  Table~\ref{variables} contains the full definitions of the new (U$_N$ and V$_N$) and old GC mass-normalised quantities.  Table~\ref{datatable2} presents the GC number specific frequencies for the galaxies studied here.

\begin{table}
 \centering
  \caption{Globular cluster specific frequencies.  S$_N$ is from Harris \& van den Bergh (1981) and T$_N$ is from Zepf \& Ashman (1993).  Subscripts of $N$, $blue$, and $red$ on the variables correspond to values calculated with the entire GC system, blue GC subpopulation and red subpopulation numbers, respectively.}\label{variables}
  \begin{tabular}{@{}rll@{}}
\hline
Variable & Definition & Notes\\
  \hline \hline
S$_N$	 & $= N_{GC} 10^{0.4(M_V + 15)}$ & $M_V$ is galaxy total V-band mag.\\
T$_N$	 & $= $ ${N_{GC}}\over{M_{\ast}/10^{9}}$  & $M_{\ast}$ is galaxy stellar mass.\\
U$_N$	 & $= $ ${N_{GC}}\over{M_{baryon}/10^{9}} $ & $M_{baryon}$ is galaxy baryon mass.\\
V$_N$	 & $= $ ${N_{GC}}\over{M_{halo}/10^{11}} $ & $M_{halo}$ is galaxy halo mass.\\
\hline
\end{tabular}
\end{table}

The present contribution to this data compilation, the isolated elliptical NGC~821, is shown as a ringed, magenta circle in subsequent figures.

\subsection{GC T-parameter Trends with Galaxy Stellar Mass and Environment}\label{tbluemass}

Galaxy total V-band magnitudes are traditionally employed to approximate the stellar masses ($M_{\ast}$) for deriving T-parameter values (e.g. Zepf \& Ashman 1993; RZS05).  However, optical photometry is not a perfect proxy for $M_{\ast}$.  For instance, one correlation that should be accounted for when approximating $M_{\ast}$ from photometry is the galaxy mass-metallicity trend, which may account for up to $1/3$ of the V-band T-parameter trends (e.g. RZS05).  Infrared light from filters such as the K-band, are much less sensitive to metallicity and therefore preferred for mitigating the galaxy mass-metallicity trend.  Age corrections as well as effects from dust extinction are also much smaller in the K-band relative to V (e.g. Olsen et al. 2004).  Another important advantage is that the K-band data come from a homogenous sample of the 2MASS (Jarrett et al.~2003) catalogue, rather than the popular, but heterogeneous compilation of V-band photometry from the RC3 (de Vaucouleurs et al. 1991).

The text in Appendix~A explains the adopted conversion between galaxy V and K-band photometry and stellar mass.  Briefly, total stellar masses are the sum of the stellar mass in the bulge and disk components of a galaxy (ellipticals were assumed to have no disk).  Appropriate mass-to-light ratios for assumed ages of the two components were taken from simple stellar population models (Bruzual \& Charlot~2003) and bulge-to-total ratios for a given galaxy Hubble Type are from Graham \& Worley (2007).

\begin{figure}
\resizebox{1\hsize}{!}{\includegraphics{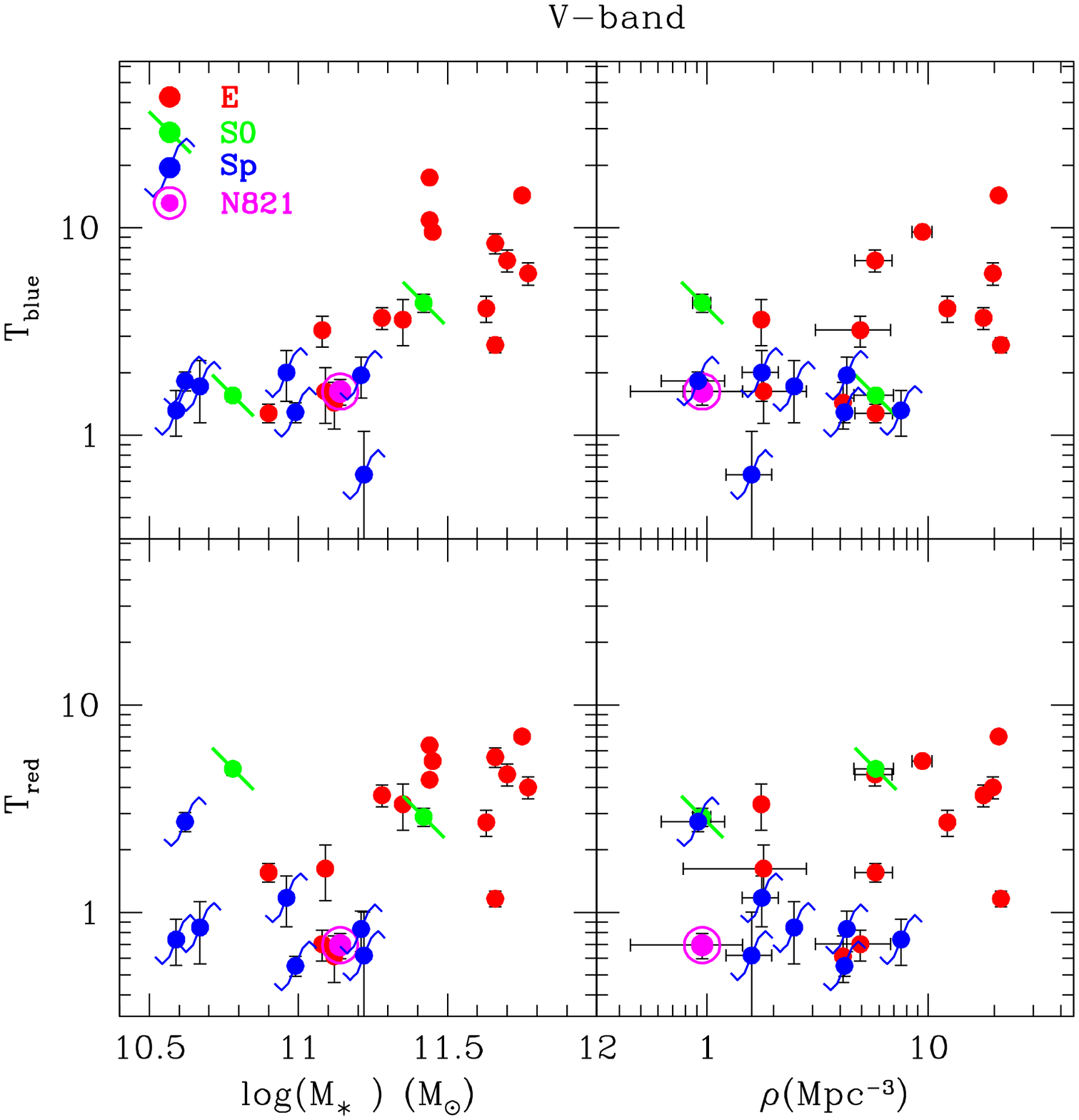}}
\caption{V-band blue and red GC T-parameters versus host galaxy stellar mass ($M_{\ast}$; {\it left panels}) and local galaxy number density ($\rho$; {\it right panels}).  T$_{blue}$ and T$_{red}$ are the number in the respective GC subpopulation normalised by its host galaxy stellar mass.  $M_{\ast}$ is computed using total V-band galaxy luminosity.  A trend for increasing T$_{blue}$ with galaxy $M_{\ast}$ is apparent, while the T$_{red}$ plot shows a weaker trend.  Galaxies with high $\rho$ values generally show larger T-parameter values.}\label{figtbluev}
\end{figure}

\begin{figure*}
\resizebox{1\hsize}{!}{\includegraphics{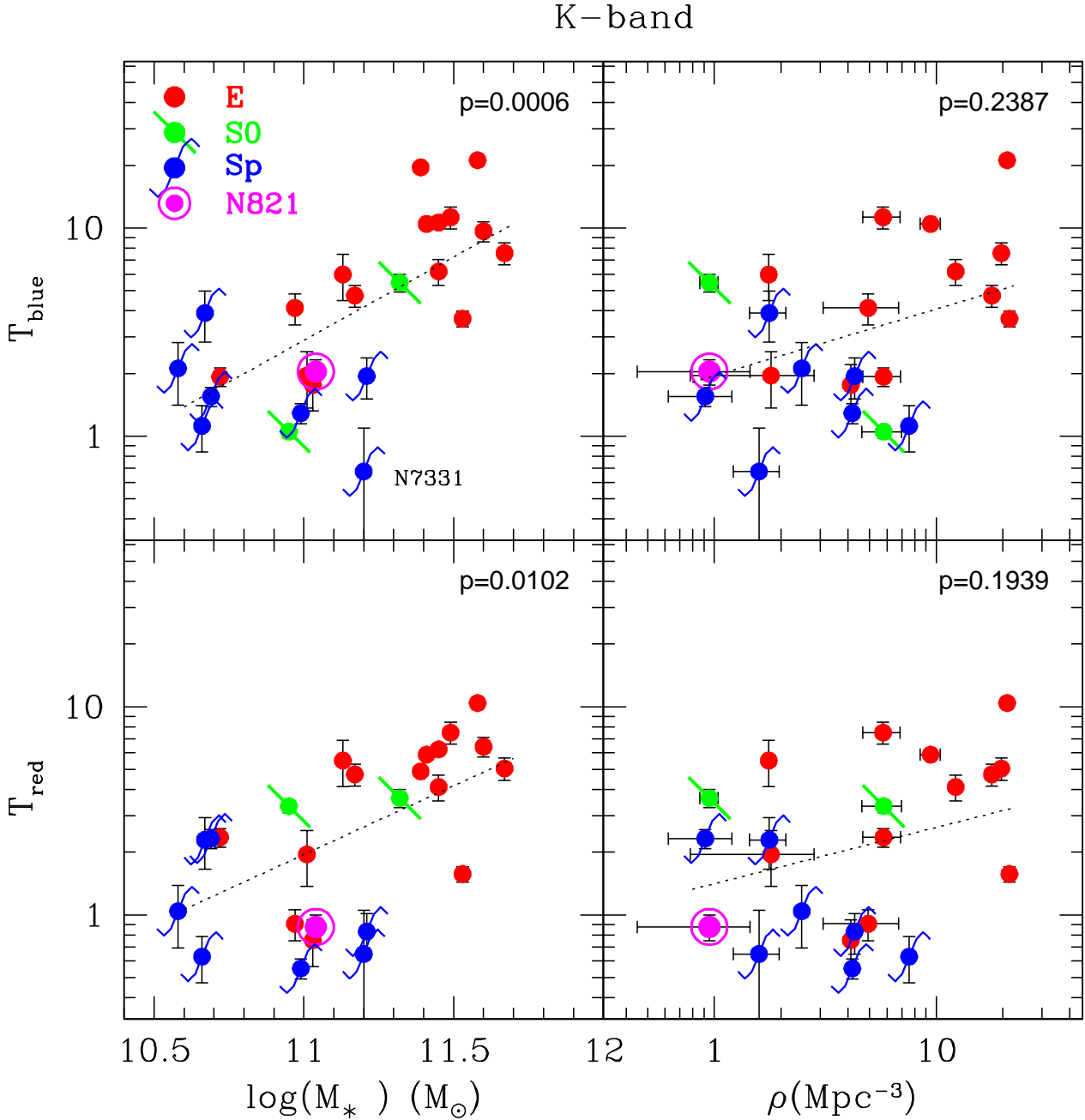}}
\caption{Same as Figure~\ref{figtbluev}, except the T-parameters and $M_{\ast}$ were computed using galaxy K-band photometry.  As detailed in the text, the p-statistics from Kendall tau correlation tests are given in the respective panels.  Dotted lines are linear fits to the data.  One outlier is evident and is labelled in the T$_{blue}$--$M_{\ast}$ plot:  the isolated spiral NGC~7331 (Rhode et al. 2007).}\label{figtbluek}
\end{figure*}

GC subpopulation T-parameter values were computed using both V and K-band $M_{\ast}$ estimates and are presented for comparison in Figures~\ref{figtbluev} and \ref{figtbluek}, respectively.  Overall, the trends are very similar in the V and K-band versions, suggesting the assumptions made to estimate $M_{\ast}$ are valid for the current sample.  Only a slight increase in apparent scatter in the K-band version of the T$_{blue}$ versus log $M_{\ast}$ plot for log $M_{\ast}<11.2$ is observed, which is likely just a small number effect.  Also, the stellar mass estimates of the Scd spiral galaxy NGC~3556 demonstrate the advantage of using K-band photometry to estimate $M_{\ast}$.  This galaxy's V-band $M_{\ast}$ estimate (log $M_{\ast}=11.0$) is 0.3 dex greater than the K-band estimate (log $M_{\ast}=10.7$), which is consistent with the idea that the V-band stellar mass is over-estimated due to a significant V-band emission from stars with ages $<5$ Gyr.  Thus, given such advantages, the K-band photometry will be used in the remaining figures and discussion regarding $M_{\ast}$.  


\subsubsection{Galaxy Stellar Mass Trends}\label{stellarmass}

Aside from one obvious outlier (see the Fig.~\ref{figtbluek} caption), it is clear that the larger sample considered here confirms RZS05, where a positive trend between T$_{blue}$ and host $M_{\ast}$ was found.  This trend implies that more massive galaxies exhibit larger stellar-mass normalised blue GC numbers.

As pointed out in RZS05, this result is inconsistent with the major-merger model of Ashman \& Zepf (1992), where the blue subpopulation of massive ellipticals is actually the sum of two blue GC subpopulations intrinsic to the two merging spiral galaxies.  In this model, to reach the high T$_{blue}$ values of massive ellipticals, the merging spiral T$_{blue}$ values must have similarly high values, contrary to what is confirmed here.  Note, Burkert, Naab \& Johansson (2007) have recently reported that their theoretical predictions for major galaxy-galaxy merger events are inconsistent with the observed central kinematic structure of massive elliptical galaxies.

RZS05 speculate the positive T$_{blue}$ trend may relate to the timescales of GC formation in a simply biased galaxy formation model (Santos 2003).  Under such a model, massive galaxy potentials hosted the earliest sites of GC formation, thus more massive galaxies may have had more time to produce blue GCs (see discussion in RZS05; Santos 2003; also Brodie \& Strader 2006).

Noticeably absent from the RZS05 T$_{blue}$ study (due to a scarcity of good total GC population estimates at the time) were massive central cluster galaxies (e.g. Virgo Cluster M87 and Fornax Cluster NGC~1399), which have been shown to host enhanced galaxy luminosity-normalised GC numbers relative to other galaxies (e.g. Harris \& van den Bergh 1981).  Included here are four galaxies located at the centre of the three nearest galaxy clusters:  Virgo, Fornax and Antlia (NGC~3258 and NGC~3268 are each located in a central sub-group within the Antlia cluster).  These galaxies show the largest T-parameter values of the sample (T$_{blue}>10$ and T$_{red}>6$), confirming previous observations.  The logarithmic scale of the T-parameter values in the plot helps support the idea that these GC systems are not actually unique, but simply at the extreme end of a general trend.

The T$_{red}$ data shows evidence for an positive analogous trend with $M_{\ast}$, but with more scatter.  RZS05 mentioned, but did not discuss the T$_{red}$ trend.  Again, this appears to imply that massive galaxies formed relatively more red GCs compared to lower mass galaxies.  Larger scatter may indicate the formation process of red GCs was subject to more stochastic variations than the processes responsible for blue GC subpopulations.  In contrast to the early-type galaxies, the T$_{red}$ values for spiral galaxies show no significant correlation with galaxy mass or $\rho$.  This may suggest the red GC subpopulations of spirals were formed in a different manner than those of early-type galaxies (e.g. effects of secular evolution vs. merger-driven spheroid formation).

As mentioned previously, the subpopulation T-parameter values of NGC~821 are fairly consistent with those of similar-mass spiral and elliptical galaxies.

Least-square linear fits were performed on the data (clipping appropriate outliers) and are presented in Figure~\ref{figtbluek} and Table~\ref{tabfits}.  Non-parametric, Kendall tau rank correlation coefficients were calculated for the data and their associated p-statistics are presented in the respective panels of Figure~\ref{figtbluek}.  The p-statistic is the probability of the null hypothesis that the data are not correlated.  According to the Kendall measure, both $M_{\ast}$ trends show a significant trend at a $\ge99.99\%$ confidence level.

\begin{table}
 \centering
  \caption{Linear fits to the T-parameters versus stellar mass and local density data presented in Figure~\ref{figtbluek}.  The p-statistic is from a non-parametric Kendall tau test for the significant of a linear correlation.}\label{tabfits}
  \begin{tabular}{@{}cccc@{}}
\hline
GC &  \multicolumn{2}{c}{log T$ $=$\alpha$ log $M_{\ast}+\beta$} & Kendall tau\\
subpop.  & $\alpha$ & $\beta$   & p-stat.   \\
  \hline \hline
Blue &$0.80\pm0.17$ & $-8.3\pm1.9$ &0.0006 \\
Red & $0.66\pm0.22$ & $-7.0\pm2.5$ &0.0102\\
\hline
GC &  \multicolumn{2}{c}{log T$ $=$\alpha$ log $\rho+\beta$} & Kendall tau \\
subpop. & $\alpha$ & $\beta$    & p-stat.  \\
  \hline \hline
Blue &$0.32\pm0.17$ & $0.29\pm0.14$ & 0.2387 \\
Red & $0.27\pm0.19$ & $0.15\pm0.15$ & 0.1939\\
\hline
\end{tabular}
\end{table}

\subsubsection{Local Galaxy Environment Trends}\label{environ1}

On the right panels of Figure~\ref{figtbluek}, the GC subpopulation T-parameters are plotted against the local galaxy number density, $\rho$, which is a proxy for the local, underlying mass-density.  While $\rho$ is not the only or even the most sensitive parameter for gauging environmental influence, it is useful for nearby galaxies and indeed provides a quantitative measure of local environmental density (e.g. Blanton \& Berlind 2007).

Compared to the T-parameter plots against mass, the $\rho$ plots in Figure~\ref{figtbluek} only hint at a positive trend.   Supporting this qualitative assessment of the data, a Kendal tau test indicates $\rho$ correlations at the $\sim99.8\%$ confidence level and the linear fits in Table~\ref{tabfits} are only significant at the $1-2\sigma$ level.  Nevertheless, on average, the T-parameters are lower in low-density environments and larger for higher values of $\rho$, with galaxies in a given environment spanning a factor of 10~in T-parameter values.  The isolated elliptical NGC~821 is among the lowest T-parameter values in the sample.

It is perhaps reasonable to ask if either $M_{\ast}$ or $\rho$ represents the dominant physical quantity effecting the T-parameter values.  As massive galaxies naturally tend to be found in higher-density environments, it is unlikely that completely separating galaxy mass and environmental effects on GC system properties is possible or even an appropriate course of action.  Nevertheless, we can consider galaxies of a similar mass or environment and ask whether or not the observed scatter in GC numbers is correlating with the third parameter.  The present sample bias towards high $M_{\ast}$ galaxies in dense environments (see~Fig.~\ref{figbias}) unfortunately makes any such inference very uncertain without a more complete dataset.  

\begin{figure}
\resizebox{1\hsize}{!}{\includegraphics{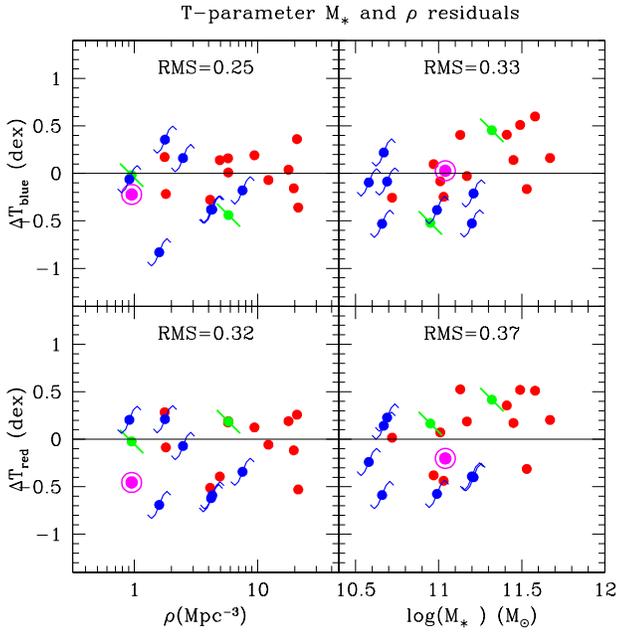}}
\caption{Residuals for the T$_{blue}$ ({\it top panels}) and T$^{}_{red}$ ({\it bottom panels}) trends versus galaxy stellar mass ($M_{\ast}$; {\it right panels}) and local galaxy density ($\rho$; {\it left panels}).  Residuals are from linear fits presented in Figure~\ref{figtbluek}.  Residual RMS values are given.  Each panel corresponds to the same panel of Figure~\ref{figtbluek}, with the third parameter not used in the fit presented in this figure on the X-axes.  Only a hint of a trend is visible in the right panels, which would indicate $M_{\ast}$ represents the dominant physical property effecting T-parameter values.}\label{figtblueres}
\end{figure}

Despite this caveat, each panel in Figure~\ref{figtblueres} shows the residuals of linear fits to the data in the corresponding panels of Figure~\ref{figtbluek}.  For instance, the residuals fit to the T$_{blue}$-$M_{\ast}$ data in the upper left panel of Figure~\ref{figtbluek}, are presented in the corresponding panel of Figure~\ref{figtblueres} against the third parameter, $\rho$.  A very shallow trend is visible in the right panels of Figure~\ref{figtblueres}, perhaps suggesting that a $M_{\ast}$ trend remains after environmental effects are removed.  This seems to suggest a stronger dependence on $M_{\ast}$.  However, this may only reflect the poor, $2\sigma$ fit of the T-parameter and $\rho$ data.  

Thus tentative evidence is found that the T-parameters depend more on $M_{\ast}$ than $\rho$, although separating the two effects is indeed challenging at the present.  It is noted that the isolated elliptical NGC~821 shows a consistent GC number to other similar-mass galaxies in higher density environments (see Figure~\ref{figtbluek}).  Other measures of local environmental density (e.g. group luminosity, group-centric distances) might help distinguish whether host galaxy mass or local environment has a stronger effect on GC system properties.  Further discussion of the environmental trends is provided in \S\ref{environ2}.

\subsection{Galaxy Baryonic Mass Trends}\label{othermass}

Before interpreting the T-parameter trends in terms of GC formation efficiency, the normalising quantity in the T-parameters, $M_{\ast}$, must be understood physically.  Total luminosities of galaxies bare a direct connection with the stellar mass content of a galaxy, once the stellar populations are understood.  However, processes such as strangulation can lead to a decreased field star formation efficiency in cluster galaxies while having no effect on GC numbers.  This effect can induce or strengthen a positive T-parameter trend with host stellar mass, thus normalising GC numbers with a total {\it baryon} mass and computing U$_N$ (see Tab.~\ref{variables}) may be more relevant if a connection between GC and galaxy formation is to be attempted.

\begin{figure}
\resizebox{1\hsize}{!}{\includegraphics{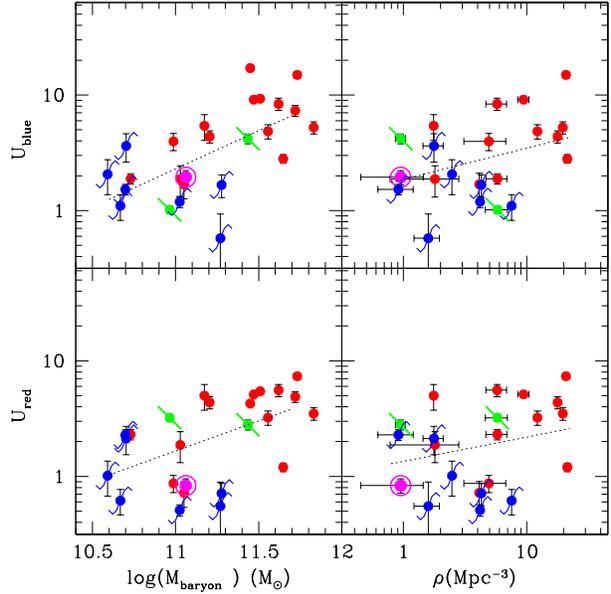}}
\caption{Similar to Figure~\ref{figtbluev}, except the U-parameter is plotted against stellar plus X-ray hot gas mass ($M_{baryon}$).  The U-parameter trends show only a slightly shallower gradient compared to the T-parameter data, thus the overall effect of including hot gas is minor.}\label{figtmass}
\end{figure}

McLaughlin (1999) explored this idea and found evidence for a universal fraction of a galaxy's baryon in its GC system (which would roughly correspond to constant U$_{blue}+$U$_{red}$ values), across a range of galaxy luminosities.  Since the McLaughlin (1999) study, many of the GC system number estimates have decreased (e.g. Rhode et al. 2007), thus an updated analysis is pertinent and is presented in Figure~\ref{figtmass} and Table~\ref{tabfitst}.  U-parameters in this figure use a galaxy mass estimate that includes both stellar masses from K-band photometry and the X-ray hot gas mass fractions from McLaughlin (1999), which is reproduced here:

\begin{equation} M_{gas} = 0.55 M_{\ast} ({L_{V}\over{10^{11} L_{\odot}}})^{1.5} \end{equation} where $L_V$ is the V-band luminosity of the host galaxy.  As galaxies show a fairly large intrinsic scatter in their X-ray luminosities at a given stellar mass (e.g. Mathews et al. 2006), the subsequent baryon mass analysis based upon Equation~2 is only useful for studying overall trends.

Interestingly, the U-parameter trends in Figure~\ref{figtmass} persist with only a slight decrease in the T-parameters gradients in Figure~\ref{figtbluek}.  The slightly shallower trends are caused by a greater hot gas fraction in massive galaxies.  Linear fits in Figure~\ref{figtmass} and Table~\ref{tabfitst} support this qualitative assessment.

\begin{table}
 \centering
  \caption{Like Table~\ref{tabfits}, except the fits are to the $M_{baryon}$ U-parameters, presented in Fig.~\ref{figtbluek}.}\label{tabfitst}
  \begin{tabular}{@{}cccc@{}}
\hline
GC &  \multicolumn{2}{c}{log U$ $=$\alpha$ log $M_{baryon}+\beta$} & Kendall tau\\
subpop.  & $\alpha$ & $\beta$   & p-stat.   \\
  \hline \hline
Blue &$0.66\pm0.17$ & $-6.9\pm1.9$ &0.0023 \\
Red & $0.52\pm0.22$ & $-5.5\pm2.4$ &0.0200\\
\hline
GC &  \multicolumn{2}{c}{log U$ $=$\alpha$ log $\rho+\beta$} & Kendall tau \\
subpop. & $\alpha$ & $\beta$    & p-stat.  \\
  \hline \hline
Blue &$0.27\pm0.15$ & $0.27\pm0.12$ & 0.2381 \\
Red & $0.21\pm0.18$ & $0.13\pm0.14$ & 0.2155\\
\hline
\end{tabular}
\end{table}

\begin{figure}
\resizebox{1\hsize}{!}{\includegraphics{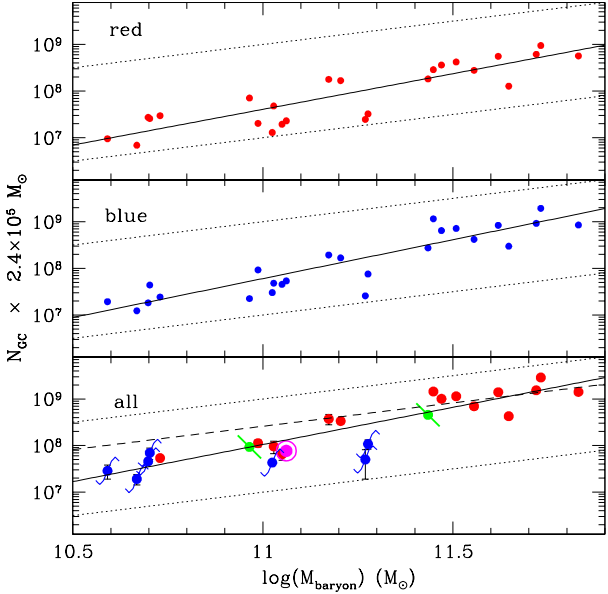}}
\caption{GC mass ($M_{GC}$) against host galaxy baryon mass ($M_{baryon}$).  Red, blue and all GCs of a given galaxy are presented in the top, middle and bottom panels, respectively.  The dotted lines correspond to constant values of $\epsilon^{b}=M_{GC}/M_{baryon}=0.01$ and $=0.0001$.  Solid lines are least-squares fits to the GC data and the dotted line in the lower panel corresponds to the ``universal'' $\epsilon^{b}=0.0026$ determined by McLaughlin (1999).  Newer GC number estimates have decreased due to better account of contamination, thus the data presented here is inconsistent with a universal GC to galaxy baryon mass fraction as found by McLaughlin (1999).}\label{figgcmassfrac}
\end{figure}

The universal GC mass ($M_{GC}$) fraction of the host baryon mass ($M_{baryon}$) found by McLaughlin (1999) was $\epsilon^{b}=0.0026$, where $\epsilon^{b}=M_{GC}/M_{baryon}$.  The bottom panel of Figure~\ref{figgcmassfrac} shows the total mass of GCs against the host baryon mass, where the GC system masses are the product of the GC population number and the average GC mass of $2.4\times10^{5} M_{\odot}$ used in McLaughlin (1999).  Included in the Figure are lines of constant $\epsilon^{b}$ and the McLaughlin (1999) universal value of $\epsilon^{b}=0.0026$.

In the present sample, $\epsilon^{b}$ of massive galaxies are consistent with the universal McLaughlin (1999) value.  However, for lower mass galaxies, the GC masses clearly show lower values than the ``universal'' $\epsilon^{b}$.  A least-squares fit to the data yields the equation log $M_{GC}=1.59\pm0.16$ log $M_{baryon}-9.5\pm1.8$, which is inconsistent with a constant $\epsilon^{b}$ (i.e. slope of 1) at the $\sim4\sigma$ level.  Both GC subpopulations show similar linear trends with only a zeropoint offset.

To reach a constant value of $\epsilon^{b}$, a factor of 5 more baryons would need to be added to the mass budget of massive ellipticals.  Otherwise, the baryonic masses of low-mass galaxies could be over-estimated by a similar factor.  Both extremes seem unlikely.  As this analysis does not include dust or gas too cold for detection in X-rays, it is technically incomplete and a more perfect analysis would involve a detailed case-by-case accounting of an individual galaxy's baryon content.  However, these contributions should be relatively minor for massive galaxies and an increase in the baryon content of low-mass galaxies would only strengthen the U-parameter trends observed here.

\begin{figure}
\resizebox{1\hsize}{!}{\includegraphics{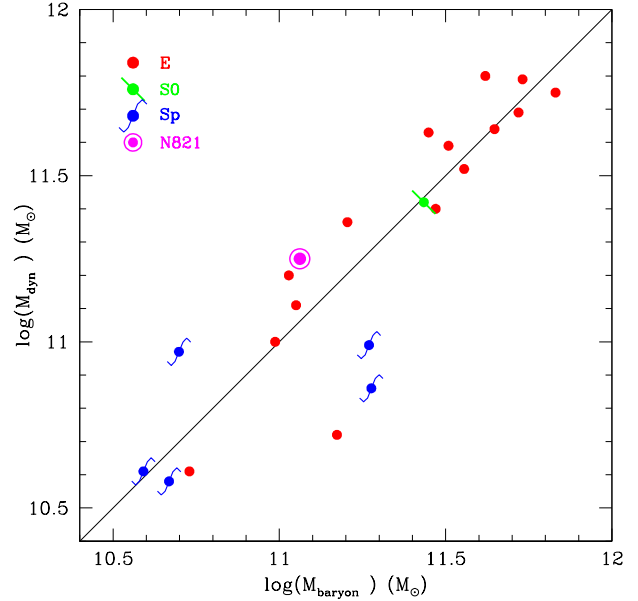}}
\caption{Comparison between dynamical and total baryon galaxy masses, whose derivation is discussed in the text.  Comparison to the 1:1 line indicates the two mass estimators yield consistent results.}\label{figstmass}
\end{figure}

This result can be corroborated with the use of dynamical masses ($M_{dyn}$).  Galaxy $M_{dyn}$ estimates are from the Cappellari et al. (2006) form of the viral theorem is used, with central velocity dispersions are from LEDA (Paturel et al. 2003) and K-band effective radii from 2MASS 20 mag arcsec$^{-2}$ isophotal radii, converted to effective radii with an empirical relationship from data in Jarrett et al.~(2003).  Figure~\ref{figstmass} shows that the dynamical masses are comparable to $M_{baryon}$, suggesting that the baryon masses derived from photometry are a good indicator of the total baryonic content.

Hence, contrary to the conclusions of McLaughlin (1999), this analysis suggests GC formation efficiency {\it increases} with the host baryon mass.  The main cause for the new interpretation is that new GC number estimates have become available, which were derived using improved contamination estimates and better constrained radial profiles of the GC systems (see e.g. Rhode et al. 2007).

\subsection{Galaxy Total Halo Mass Trends}\label{halomass}

\begin{figure*}
\resizebox{1\hsize}{!}{\includegraphics{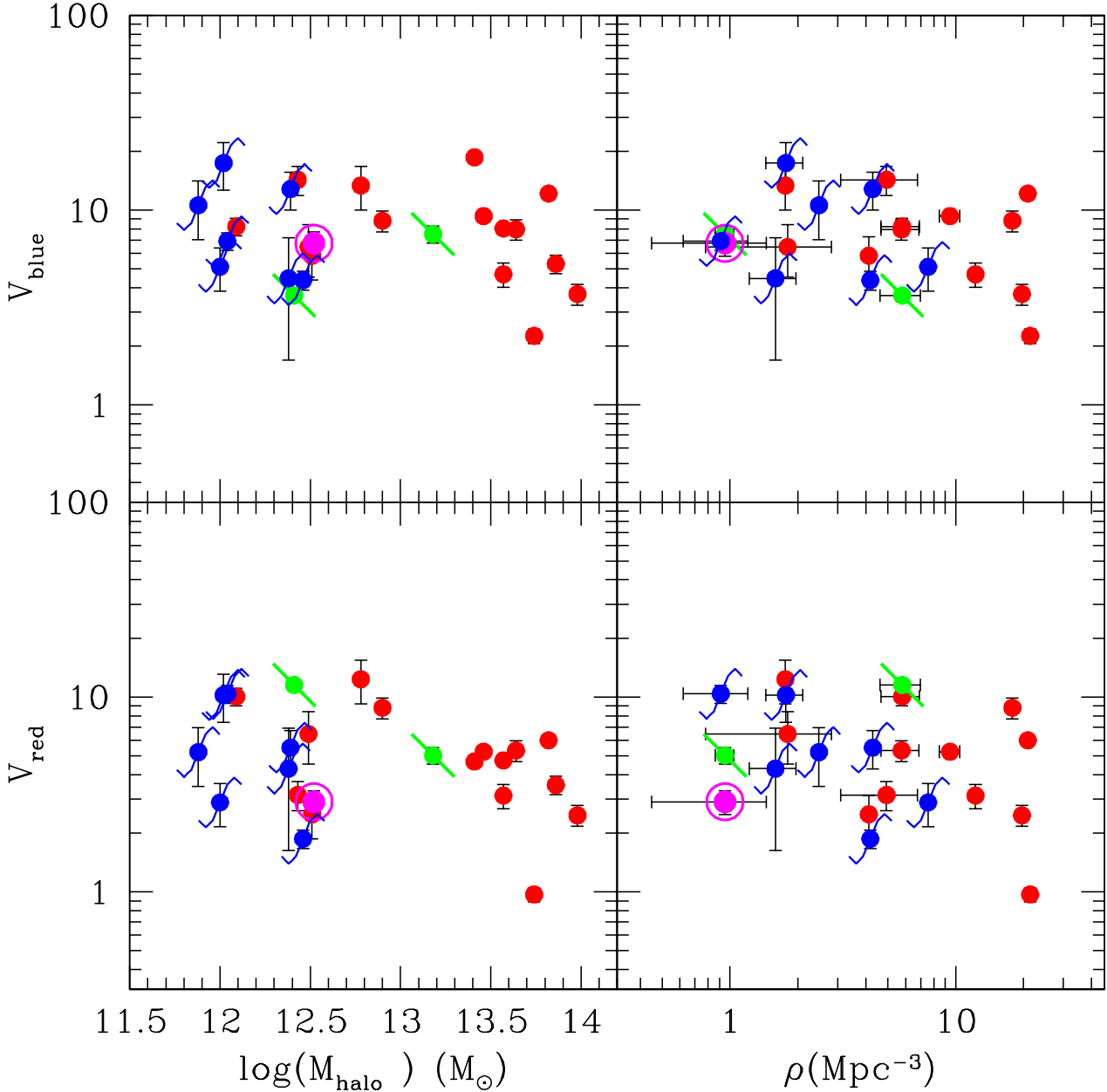}}
\caption{GC halo mass normalised numbers (V$_N$) against galaxy halo mass ($M_{halo}$), which is derived from the weak lensing analysis of Mandelbaum et al. (2006).  V-parameter values are normalised to a halo mass of log $M_{halo}=11$ and show no significant correlation with either $M_{halo}$ or $\rho$.  This suggests GCs formed in direct proportion to the total halo mass of a galaxy.  See text for detailed discussion.}\label{fighmass}
\end{figure*}

A final quantity worth exploring for the GC number normalisation mass is galaxy total halo mass, $M_{halo}$.  Mandelbaum et al. (2006) performed a galaxy-galaxy weak lensing analysis of SDSS galaxies and derived mean halo masses for galaxies spanning a range of stellar masses (log $M_{\ast}=9.9$ to $=11.6$).  They found a non-linear, positive gradient between galaxy stellar and halo mass that steepens significantly for galaxies with log $M_{\ast}>11$.  Although such pioneering work must be used with caution, the Mandelbaum relationship can be used to derive total halo masses and hence new normalised GC numbers, the V-parameters (see Tab.~\ref{variables}), calculated.  

A few caveats should be considered first.  As the lensing sample of Mandelbaum et al. (2006) occupy a range of redshifts ($0.02 < z < 0.35$), the halo masses might be underestimated for the local galaxy sample, because halos (and their GC systems) may grow with time.  However, Parker et al. (2007) find little or no evidence for halo mass growth over the redshift range $0.2 < z < 0.8$, and if this trend can be extended to the present epoch, would suggest the Mandelbaum et al. (2006) halo masses are roughly valid for nearby galaxies.  Another potential concern is that halos likely show intrinsic variations for a given galaxy stellar mass, thus the halo masses assumed may not be accurate representations of the specific galaxies considered here.  Mandelbaum et al. (2006) must stack their sample of galaxies to detect a lensing signal.  Their work therefore gives a good average halo in a statistical sense, but will not apply to an individual galaxy.  Mathews et al. (2006) provide some evidence this variation is fairly significant for a given galaxy $M_{\ast}$, thus the subsequent analysis is only valid for overall trends.  As to the robustness of the Mandelbaum et al. halo masses, it is noted that other studies have produced similar results.  Parker et al. (2007) confirm the Mandelbaum et al. (2006) halo masses for L$^{\ast}$ galaxies with a deeper weak-lensing analysis.  Halo occupation statistical modelling by van den Bosch et al. (2007) predicts a similar scaling of halo and central stellar mass.

\begin{figure}
\resizebox{1\hsize}{!}{\includegraphics{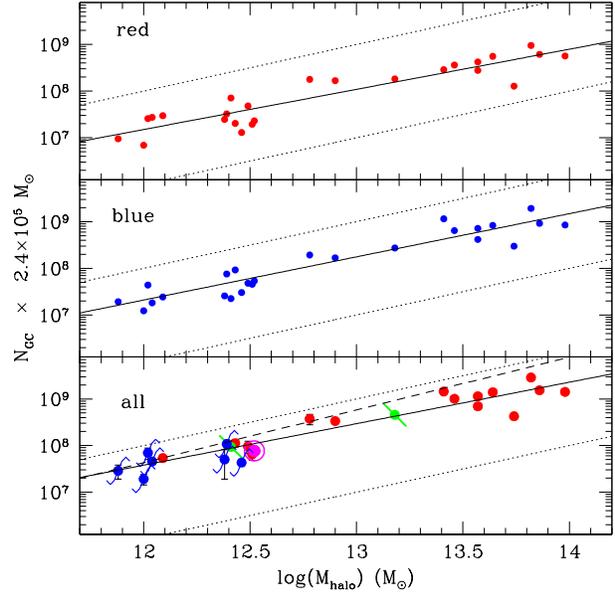}}
\caption{Total mass in GCs ($M_{GC}$) against host galaxy halo mass ($M_{halo}$).  The layout is as in Figure~\ref{figgcmassfrac} with data from Figure~\ref{fighmass}.  Dotted lines correspond to arbitrary constant values of $M_{GC}/M_{halo}=\epsilon^{h}=1\times10^{-4}$ and $=1\times10^{-6}$.  The data are consistent with constant GC mass fractions.  Mean mass fractions of blue, red and all GCs are $\epsilon^{h}= 2.0\times10^{-5}, 1.3\times10^{-5},$ and $3.2\times10^{-5}$, respectively.  Dashed line is the extrapolated prediction from theoretical analysis of Kravtsov \& Gnedin (2005).}\label{figgcmassfrachalo}
\end{figure}

The V-parameter values are arbitrarily normalised to a halo mass of log $M_{halo}=11$ (see Tab.~\ref{variables}) and are presented in Figure~\ref{fighmass}.  Mandelbaum et al. (2006) relationships for early and late-type galaxies were applied to the appropriate galaxy type for galaxies log $M_{\ast}>11$.  Below this stellar mass, Mandelbaum et al. find no significant difference between the halo masses of early and late-type galaxies, and so the early-type relation was used in this range.  Figure~\ref{fighmass} shows a striking contrast to the T and U-parameter plots discussed previously:  the positive trend with galaxy mass is gone, suggesting GCs (of both subpopulations) formed in a constant proportion to the host halo mass, irrespective of environment or host galaxy type.  The $M_{halo}$ data are presented in alternate form in Figure~\ref{figgcmassfrachalo} and are statistically consistent (according to least-squares fits) with a constant GC mass fraction of the host halo mass.  The best-fitted line to the total GC system properties is:  log $M_{GC}=0.94\pm0.06-3.77\pm0.76$.

The idea that GCs formed in direct proportion to the mass of its host halo is attractive for a number of reasons (see discussion in Blakeslee, Tonry \& Metzger 1997 and Blakeslee 1999).  Such a model gels with observational evidence that most GCs formed in the very early ($>10$ Gyr) Universe, conceivably before galaxy star formation histories began to be influenced by processes like AGN feedback or strangulation.  In this framework, central massive galaxies such as NGC~4486 or NGC~1399, show ``enhanced'' GC populations (see \S\ref{stellarmass}) because they are associated with an extremely massive dark matter halos.  A useful consequence is that a direct relationship between GC system numbers and total halo mass could be derived.  Such a relationship would help characterise the undetected or removed baryon content of a galaxy.

A few theoretical predictions exist for this analysis (Santos 2003; Diemand et al. 2005; Kravtsov \& Gnedin 2005).  In \S\ref{reion} below, the blue GC results will be compared to the Diemand et al. (2005) predictions for a variable reionisation epoch.  Santos (2003; see also RZS05 and Rhode et al. 2007) predicted that galaxy ``biasing'' will naturally lead to greater blue GC numbers in more massive galaxies (contrary to what is found here) assuming blue GC formation is universal truncated by a mechanism like reionisation.  Santos (2003) suggested the sites of massive galaxies were biased to host the first GC formation regions, thus these galaxies would have had a longer opportunity to produce blue GCs before a universal truncation.  

Kravtsov \& Gnedin (2005) reported a correlation between the mass of a GC system and its host halo mass for simulated halos between log $M_{halo}=10.8$ and $=11.4$.  Although their analysis is not valid for the halo mass range considered here, Figure~\ref{figgcmassfrachalo} shows the correlation extrapolated from lower masses.  Kravtsov \& Gnedin (2005) point out that the Milky Way (here with log $M_{halo}=12.5$) falls on their extrapolated relation.  Supporting this observation, the present analysis suggests all galaxies in the halo mass range log $M_{halo}~12$ to $~13$ are roughly consistent with their extrapolated predictions.  Only the most massive galaxies appear to deviate.

\subsection{Comparison between Galaxy Halo and Baryon Mass Trends}\label{halobaryonmass}

Although galaxy stellar to halo mass fractions ($M_{\ast}/M_{halo}$) vary with host galaxy mass (e.g. Mandelbaum et al. 2006), on scales of galaxies the baryon fraction might roughly resemble the universal baryon fraction (i.e. $M_{baryon}/M_{halo}\approx\Omega_{b}/\Omega_{m}\approx0.17$; WMAP3, Spergel et al. 2007).  If a universal baryon fraction is applied to the $M_{GC}$-$M_{baryon}$ data in Figure~\ref{figgcmassfrac}, to make a $M_{GC}$-$M_{halo}$ plot analogous to Figure~\ref{figgcmassfrachalo}, only an X-axis shift to larger halo masses would occur and the form of the $M_{GC}/M_{baryon}$ trend will directly translate into to a $M_{GC}/M_{halo}$ trend.  This apparently contradicts the roughly constant $M_{GC}/M_{halo}$ shown already in Figure~\ref{figgcmassfrachalo}.  A possible explanation for this ``discrepancy'' is that the local baryon fraction decreases with increasing host halo mass.  Although this idea is interesting, a few analytical checks and further discussion are needed before making any conclusions.

It is first worth reviewing the empirical relationships used to derive the $M_{baryon}$ and $M_{halo}$ values.  Total galaxy baryon masses are from the sum of total stellar mass (from K-band photometry; see \S\ref{datacat}) plus an X-ray hot gas component (between $40\%$ and 0\% of $M_{\ast}$ for high and low-mass galaxies, respectively; see Eq.~2~in \S\ref{othermass} and McLaughlin 1999).  The hot gas relation is normalised to NGC~4486 (M87), for which McLaughlin (1999) estimate $\sim40\%$ of the baryons are in hot gas within 100 kpc of the galaxy centre.  To make $M_{GC}/M_{baryon}\sim$ constant, this percentage would have to increase to an unlikely value of $\sim90\%$ (i.e. thereby increasing baryons by a factor of five, see \S\ref{othermass}), suggesting the current baryon calculations are reasonable.  It was also found that the $M_{baryon}$ are consistent with mass estimates inferred from central galaxy dynamics (see Fig.~\ref{figstmass}).

The contradiction may also be alleviated if the Mandelbaum et al. (2006) halo masses are under-estimated for low-mass galaxies or over-estimated for high-mass galaxies.  Interestingly, Mandelbaum et al. (2006) find that for log $M_{\ast}<11$, early and late-type galaxies (classified as such by their light profiles) have similar $M_{halo}$ values, on average, but for log $M_{\ast}>11$ early-type galaxies are hosted by significantly more massive halos than those surrounding late-type galaxies.  Mandelbaum et al. point out that massive early-type galaxies tend to be found at the centres of galaxy groups or clusters and speculate the lensing masses of massive early-type galaxies actually reflect group/cluster mass potentials rather than galaxy-sized halos.  If this interpretation of Mandelbaum et al. is correct, then the $M_{halo}$ values of massive early-type galaxies in Figure~\ref{fighmass} and \ref{figgcmassfrachalo} may not be comparable to those galaxies of log $M_{\ast}<11$.  Applying the late-type trend (i.e. from non-central galaxies) trend to all galaxies of log $M_{\ast}>11$ increases their halo V-parameter values by an amount that may remove the discrepancy with the $M_{baryon}$ U-parameter trends.  This seems to support the Mandelbaum et al. interpretation of the observed halo mass difference between early and late-type galaxies.

On the other hand, four of the galaxies considered here are actually located at the centres of massive groups or galaxy clusters.  It is also very likely that very massive, non-central ellipticals (e.g. NGC~4472 and NGC~3268) must have once been at the centre of a large galaxy group, which has since been incorporated into a galaxy cluster.  As these massive ellipticals host enhanced GC systems for their baryonic masses (see \S\ref{othermass}), it is worth asking whether there exists a connection between the GC system and the larger cluster-sized halos they are (or once were) at the centres of.  Blakeslee et al. (1997; see also Blakeslee 1999) favoured such a connection after they found that the S$_N$ values of massive elliptical galaxies at the centres of clusters correlate strongly with the host cluster mass.

To remove the apparent discrepancy between the U$_N$ and V$_N$ values, the U$_N$ values of the massive galaxies must be brought down to the observed levels for ``normal'' low-mass galaxies, thus more baryons need to be added to the central galaxy's mass budgets.  Assuming massive elliptical GC systems are or once were associated with a cluster-sized halo, a more appropriate M$_{baryon}$ estimate might include {\it baryons associated with the intra-cluster medium}.  The contribution from an intra-cluster baryon component for a central cluster galaxy mass budget can be crudely estimated here.  Gonzalez, Zaritsky \& Zabludoff (2007) found that the luminosity of the intra-cluster light is roughly greater than 5 times the light in the brightest cluster galaxy, so including this component in the cluster baryon budget would add roughly a factor of $\sim5$ or more baryons.  This factor is very similar to the increase in baryon mass needed in the most massive ellipticals to produce a constant value of U$_N$ for all galaxies (see \S\ref{othermass}).  Baryons from intra-cluster gas will push this factor beyond five, which might be counter-balanced by allowing for a GC mass component from intra-cluster GCs.  Though a future, more detailed investigation is needed, this analysis lends some support to the idea that central galaxy GC system numbers are indeed proportional to total cluster mass as found in Blakeslee et al. (1997) and Blakeslee (1999).  Furthermore, the inclusion of baryons from the intra-cluster medium surrounding central galaxies may remove some, if not all, the discrepancy between the U$_N$ and V$_N$ trends (implying the baryon fraction, on the scales of galaxy clusters, is equivalent to the universal baryon fraction).  This again suggests that GCs formed in proportion to the total halo mass available, be it from a cluster or galaxy-sized halo.  It is therefore appropriate to use the Mandelbaum et al. (2006) early-type (i.e. central) relationship on these massive elliptical galaxies, as in Figures~\ref{fighmass} and \ref{figgcmassfrachalo}.

Clearly further work is needed on many fronts.  GC system numbers show at least some promise as indicators of halo mass, although individual halo estimates of galaxies in the local Universe (e.g. from GC dynamics, Romanowsky 2006) are needed to confirm and better quantify such a correlation.  Carefully determining the baryon content of individual galaxies and comparing their GC system numbers to other galaxies will help determine the extent of the possible GC number and baryon/halo mass discrepancy and whether the local baryon fraction varies with host galaxy mass.  GC systems of dwarf galaxies are very interesting systems to add to this analysis because they are thought to be dominated by dark matter, thus they might host many more GCs than otherwise would be predicted from their stellar masses.  

\subsection{Constraining Reionisation}\label{reion}

Although contrary theories exist (e.g. Cen 2001), several workers have speculated how GC metallicity bimodality can result from the truncation of GC formation during cosmic reionisation (e.g. Santos 2003; Beasley et al. 2002; Moore et al. 2006).  In this picture, metal-poor GC formation halted when reionisation increased the ambient temperature beyond the threshold for efficient star formation.  Some period of time later, after gas had become enriched with metals and cooled, a second GC formation epoch commenced resulting in the second GC metallicity subpopulation.  Though the detailed physics of GC formation is not well-constrained, GCs provide ideal tracers of dark matter from very early stages of galaxy formation (Diemand et al. 2005), thus they could provide observational constraints on the nature of reionisation (Santos 2003; Diemand et al. 2005; Moore et al. 2006; Bekki 2005 and for intra-cluster GCs see Bekki \& Yahagi 2006 and Power, Drinkwater \& Silk 2007).

Diemand et al. (2005) predicted that galaxies with the largest V$_{blue}$ values will tend to reside in regions where reionisation occurred later:  i.e. locations where metal-poor GC formation was more prolonged.  In the context of this model, the constant V$_{blue}$ values for all galaxies (see~$\S\ref{halomass}$) suggests the epoch was roughly coeval throughout the Universe.  A flat trend with environmental density may also imply that reionisation was not significantly biased to start in a particular type of environment (e.g. Weinmann et al. 2007).  Detailed modelling of the range of galaxy halo masses examined here is required before the observations can provide tighter constraints on the exact epoch of reionisation and the nature of its propagation throughout the Universe.

\subsection{Globular Cluster Systems and their Environment}\label{environ2}

Brodie \& Strader (2006) discussed recent observational evidence that may imply GCs formed in the proto-galactic fragments before galaxy assembly began.  Under such a scenario, GC numbers may provide a measurable proxy for the initial conditions of a given galaxy (e.g. the total available mass or environmental conditions).  Thus GC numbers could serve as a basis quantity for comparison between different galaxies to isolate evolutionary processes that shape galaxies into their present-day form.  Before GC systems can be used as such, the fundamental relationship between their own properties and the initial conditions (e.g. available mass and their environment at formation), must be understood independent of their host galaxies.  The objective of this section is to determine whether or not GC system numbers (i.e. GC formation efficiencies) show intrinsic differences with the local environmental density, independent of GC system number trends with the host galaxy mass.  Galaxy baryon masses are preferred over M$_{\ast}$ for the normalising quantity in the GC specific frequencies, because M$_{baryon}$ (like GC numbers) is not influenced by certain environmentally-dependent processes that will effect a galaxy's total stellar mass content. 

In the past, the answer to the question whether or not GC system formation is sensitive to the local environment, has tended towards the affirmative:  GC systems were thought to be {\it intrinsically} more populous in higher density environments, implying GC formation efficiency was somehow higher in these environments (e.g. Harris 1991; West 1993; Kumai, Hashi \& Fujimoto 1993).  Analysis presented in Section~\ref{environ1} largely confirms Kumai et al. (1993) and West (1993), who found a positive trend between S$_N$ and the Tully (1998a) environmental density parameter.  However, the robustness of this result is questioned here.  Due to a bias in the early-type GC system number estimates available, both the present and previous studies (see Fig.~4~in Kumai et al. 1993; a similar bias exists in the West 1993 sample) do not have access to GC systems of low-mass cluster galaxies, thus there is not sufficient leverage to separate mass from environmental effects on the GC system numbers (see \S\ref{environ1}).

Such studies have therefore not presented solid evidence for a S$_N$ (or T$_N$ and U$_N$; see \S\ref{othermass}) trend with local environmental density once trends with galaxy mass are removed.  Nor has contrary evidence been presented, thus the extent of an environmental impact on non-central GC systems (independent to the host mass) is subject to debate.  However, the tentative result for a constant halo mass-normalised number of GCs (i.e. V$_N$) over the entire range of galaxies masses considered here, is consistent with a model where GC system numbers show no or very little environmental dependence.

Among giant ellipticals located at the centres of galaxy clusters, Blakeslee et al. (1997; see also West et al. 1995; Harris, Harris \& McLaughlin 1998) found some evidence for an increase in S$_N$ with increasing number of nearby galaxies, i.e. environmental density.  In addition, they found a strong correlation between central galaxy's S$_N$ values and the mass of the host cluster.  It is noted that there exists an empirical trend where more massive clusters tend to show larger central galaxy number density values (e.g. Bahcall 1981).  As the central galaxies studied by Blakeslee et al. (1997) occupy a very narrow range of stellar masses, a S$_N$ trend with mass or environment can be caused by a variation in the GC and/or field star formation efficiencies.  Blakeslee et al. (1997; see also Blakeslee 1997) and others (Blakeslee 1999; Harris et al. 1998; McLaughlin 1999) adopt the view that the central galaxies of more massive clusters become increasingly under-luminous (from an increased inefficiency of field star production) and therefore central GC system numbers scale with the total cluster halo mass and/or local environmental density.  It is again apparent that if environmental effects are to be isolated, baryon masses should be used instead of stellar masses, especially for massive ellipticals where hot gas contributes significantly to M$_{baryon}$.

McLaughlin (1999) performed a careful baryon mass accounting for the central Virgo (M87) and Fornax (NGC~1399) cluster galaxies and found a very similar fraction of the galaxy's baryon mass is contained in these GC systems.  Given that Virgo and Fornax clusters exhibit noticeably different masses and central galaxy number densities (see e.g. Jord{\'a}n et al. 2007a), the McLaughlin (1999) results provide strong evidence that GC formation efficiency in central ellipticals is remarkably similar.  Thus, while S$_N$ of massive ellipticals clearly correlates with the cluster mass and possibly with local environmental density (e.g. Blakeslee et al. 1997), it isn't clear that U$_N$ does.  Furthermore, observational evidence suggests that for massive ellipticals V$_N$ is roughly constant (Blakeslee et al. 1997; Blakeslee 1999; \S\ref{halomass}).  If the above were confirmed, GC formation efficiency is likely to be insensitive to local environmental density and hence GC system numbers can serve as a good indicator of the initial mass available to a host galaxy.

\section{Summary and Conclusions}

The present work on NGC~821 gives the best constraints on an elliptical galaxy's GC system properties located in a truly isolated environment.  All measured properties of its GC system indicate a fair resemblance to similar-mass elliptical galaxies in higher density environments.  Its isolated condition has therefore produced no distinguishing features compared to ellipticals in cluster environments.  

This result is supported by a comprehensive comparison of the GC systems in $25$ galaxies spanning a range of Hubble Types, masses and environments.  Two new GC specific frequencies are introduced:  U$_N$ and V$_N$, which refer to GC numbers normalised by the host galaxy baryon and halo masses, respectively (see Tab.~\ref{variables}).  With better galaxy mass estimates using 2MASS K-band galaxy photometry and the inclusion of an hot X-ray gas component, the new compilation suggests that the galaxy baryon mass ($M_{baryon}$) normalised number of metal-poor and metal-rich GCs (i.e. the GC subpopulation U-parameters: U$_{blue}$ and U$_{red}$) increases with host galaxy baryon mass.  This result is at odds with idea that GC formation is directly proportional to the $M_{baryon}$ of the host galaxy, as concluded in the study of McLaughlin (1999).  The different conclusion reached here is likely due to improved contamination analysis used in recent GC system studies.

Spiral galaxies in the sample generally show lower T$_{blue}$ values compared to massive ellipticals, ruling out the possibility that massive elliptical GC systems were formed via the major mergers of spiral galaxies (Ashman \& Zepf 1992).  Note, the T-parameter values of spirals are similar to low-mass ellipticals.  Also, the spiral GC system studies used here are biased towards field spirals, and do not include spirals located on the outskirts of galaxy clusters.

New weak lensing halo mass ($M_{halo}$) estimates from Mandelbaum et al. (2006) were employed to investigate GC formation efficiency relative to the host halo mass.  The V$_N$ data appears to be consistent with a constant GC number per unit halo mass, as found among massive central cluster galaxies by Blakeslee et al. (1997).  GC system numbers might therefore provide an estimate of the total halo mass of a hosting galaxy.  This analysis is compared to the U-parameter trends.  There is a possible discrepancy between the two if the local baryon fraction does not decrease with increasing galaxy mass.  Part of this problem may be removed if the GC system numbers of very massive ellipticals are directly proportional to the total halo mass of the cluster these giant ellipticals tend to be at the centres of.  Caveats and implications are discussed in \S\ref{halomass} and \S\ref{halobaryonmass}.

The T and U-parameters are also found to correlate with local galaxy environment, as traced by a modified version of the Tully (1998a) $\rho$ parameter.  Due to a bias in the present and past (Kumai et al. 1993; West 1993) galaxy samples, it is currently difficult to determine whether $M_{baryon}$ or $\rho$ has a greater influence on GC numbers.  In central cluster galaxies, while S$_N$ may correlate with environment (e.g. Blakeslee et al. 1997), U$_N$ likely does not (McLaughlin 1999), supporting a model where GC formation efficiency is insensitive to the local environment.  This model is further supported by the observation that V$_N$ values are roughly constant among massive, central ellipticals (Blakeslee et al. 1997) and the assortment of galaxies studied here (\S\ref{halomass}). 

Theoretical work by Diemand et al. (2005) provides an interesting connection between the number of blue GCs and the local epoch of reionisation.  They predict the relative number of blue GCs should be higher in regions of the Universe where reionisation took place later because the time available for blue GC formation was longer in duration.  The GC systems considered in the present work may show a constant number of blue GCs per unit halo mass, V$_{blue}$, suggesting the reionisation did not vary significantly with host halo mass or the local environmental density.\\

\noindent{\bf Acknowledgments}\\

We appreciate useful discussions with F. Abdalla, T. Mendel, F. Pearce, and R. Proctor.  A warm thanks is given to the WIYN crew for supporting us during the observations.  S. Larsen kindly allowed us to use his WFPC2 pipeline.  Excellent comments from an anonymous referee significantly improved the quality of the text.  DF thanks the Australian Research Council for financial support.  JPB and JS are supported by NSF grant AST-0507729.  JSG thanks the University of Wisconsin Graduate School for partial support of his research.  This research has made use of the NASA/IPAC Extragalactic Database (NED) which is operated by the Jet Propulsion Laboratory, California Institute of Technology, under contract with the National Aeronautics and Space Administration.

\appendix
\section[]{Data Compilation}\label{appendix}

Literature compilations of robust GC system properties and their references are presented in Table~\ref{datatable1}.  GC specific frequencies are tabulated in Table~\ref{datatable2}.  Galaxy Hubble Types were taken from NED and the local density parameter, $\rho$, was computed from the Tully (1988a) catalogue, as described in \S\ref{datacat}.  Most distances were taken from the NED1D\footnote{http://nedwww.ipac.caltech.edu/level5/NED1D/intro.html} database.  Errors on T-parameter values are from published number uncertainties.  If such numbers were unavailable, these errors reflect Poisson uncertainties of the GC numbers.  References for the distances and GC system properties are given in the Table~\ref{datatable1} caption.

To derive galaxy stellar masses from photometry, a mass-to-light ratio that varies with Hubble Type was used.  Galaxy discs show younger stellar ages compared to bulge components, thus the total measured photometry was divided into these two components with empirical K and B-band (B-band is used here for the V-band galaxy photometry) bulge-to-total ratios from Graham \& Worley (2007).  Bruzual \& Charlot (2003) theoretical mass-to-light ratios (assuming a metallicity of [Fe/H] $=+0.09$) for a young (5 Gyr) and old (12 Gyr) stellar population were employed for the disc and bulge components, respectively.  Elliptical galaxies have no disc component, thus their light is assumed to be dominated by an old stellar population.  These mass estimates only include stellar mass and not the estimated gas produced during stellar evolution.  Mass-to-light ratios for a given Hubble Type are presented in Table~\ref{tabml}.  Note the values in Table~\ref{tabml} are very similar to those used in the original definition of T$_N$ by Zepf \& Ashman 1993, suggesting their empirical mass-to-light ratios vary due to stellar population effects.

As discussed in \S\ref{tbluemass}, approximating galaxy masses with K-band photometry is preferred over V-band.  Stellar masses in Table~\ref{datatable1} are from K-band photometry, which is from the 2MASS Extended Source Catalogue (Jarrett et al. 2000) and the 2MASS Large Galaxy Atlas (Jarrett et al. 2003) when appropriate.  V-band photometry is from the RC3 (de Vaucouleurs et al. 1991).  For M31, the 2MASS K-band photometry is likely underestimated (Jarrett et al.~2003) thus the V-band total luminosity was used for the stellar mass calculations.  A total K-band estimate for the Milky Way could not be found, so the V-band mass estimate was adopted.  The V-band photometry for NGC~1399 is from McLaughlin (1999) for reasons discussed therein.  All galaxy photometry is corrected for Galactic extinction.  Host baryon mass, M$_{baryon}$, is the sum of M$_{\ast}$ and a host gas mass estimate that depends on the host V-band magnitude (derived by McLaughlin 1999; see Eq.~1 in \S\ref{othermass}).  Halo masses, M$_{halo}$, are derived using weak-lensing M$_{halo}$ estimates (Mandelbaum et al. 2006; see \S\ref{halomass}).

The GC total number for NGC~1399 was derived using the surface density profile from Bassino et al. (2006B).  Assuming the GCLF form found for NGC~1399 by Dirsch et al. (2003), the Bassino et al. (2006B) surface densities were increased by 60\% to account for incompleteness in their study.  The best-fit de~Vaucouleurs profile was integrated to the GC system edge (45\arcmin; Bassino et al. 2006B), which yielded a total GC population estimate of $5800\pm700$.

\begin{table*}
 \centering
  \caption{REFERENCES --
({\it a}) Tonry et al. 2001; Jensen et al. 2003; ({\it b}) Rhode \& Zepf 2001; ({\it c}) Cantiello et al. 2005; ({\it d}) Spitler et al. 2007, in prep.; ({\it e}) Tamura et al. 2006; ({\it f}) N. Tamura priv. comm. 2007; ({\it g}) Gomez, Richtler 2004; ({\it h}) Dirsch et al. 2003; Bassino, Richtler, \& Dirsch  2006b; present work, see Appendix A; ({\it i}) Dirsch et al. 2003; ({\it j}) Rhode \& Zepf 2004; ({\it k}) Bassino, Richtler, \& Dirsch 2008, submitted; ({\it l}) Dirsch, Schuberth, \& Richtler 2005; ({\it m}) McConnachie et al. 2005; ({\it n}) Ashman \& Zepf 1998; ({\it o}) Saha et al. 2006; ({\it p}) Rhode et al. 2007; ({\it q}) Rejkuba 2004; ({\it r}) Harris et al. 2006; ({\it s}) present work; ({\it t}) Gregg et al. 2004; ({\it u}) Forbes, Georgakakis, \& Brodie 2001; ({\it v}) Forte et al. 2001; ({\it w}) Bassino, Richtler, \& Dirsch  2006a; ({\it x}) Rhode \& Zepf 2003; ({\it y}) Mould et al. 2000
}\label{datatable1}
  \begin{tabular}{@{}llccccccccc@{}}
\hline
Galaxy & Type & m--M & M$_{K}^{T}$ & M$_{V}^{T}$ &  log($M_{\ast}$) & log($M_{baryon}$) & log($M_{halo}$) & $\rho$ &  N$_{GC}$ & Red GC \\
 &  & mag & mag & mag &  M$_\odot$ & M$_\odot$ & M$_\odot$ & Mpc$^{-3}$ &  & \%     \\
\hline\hline
N4472 & E & $30.89^{a}$ & $-25.50$ & $-22.51$ & 11.67 & 11.83 & 13.98 & $19.68\pm 0.59$ & $ 5900\pm  721^{b}$ & $40^{b}$ \\
N1407 & E & $32.00^{c}$ & $-25.32$ & $-22.26$ & 11.60 & 11.72 & 13.86 & $-$ & $ 6400\pm  700^{d}$ & $40^{d}$ \\
N4486 & E & $31.03^{a}$ & $-25.26$ & $-22.47$ & 11.58 & 11.73 & 13.82 & $20.89\pm 0.72$ & $12000\pm  800^{e}$ & $33^{f}$ \\
N4374 & E & $31.32^{a}$ & $-25.15$ & $-22.25$ & 11.53 & 11.65 & 13.74 & $21.38\pm 1.09$ & $ 1775\pm  150^{g}$ & $30^{g}$ \\
N1399 & E & $31.34^{a}$ & $-25.03$ & $-22.34$ & 11.49 & 11.62 & 13.64 & $ 5.76\pm 1.10$ & $ 5800\pm  700^{h}$ & $40^{i}$ \\
N4406 & E & $31.01^{a}$ & $-24.94$ & $-22.17$ & 11.45 & 11.56 & 13.57 & $12.25\pm 0.70$ & $ 2900\pm  415^{j}$ & $40^{j}$ \\
N3268 & E & $33.00^{c}$ & $-24.94$ & $-21.70$ & 11.45 & 11.51 & 13.57 & $-$ & $ 4750\pm  150^{k}$ & $37^{k}$ \\
N4636 & E & $31.24^{l}$ & $-24.84$ & $-21.73$ & 11.41 & 11.47 & 13.46 & $ 9.44\pm 0.97$ & $ 4200\pm  120^{l}$ & $35^{l}$ \\
N3258 & E & $33.00^{c}$ & $-24.79$ & $-21.70$ & 11.39 & 11.45 & 13.41 & $-$ & $ 6000\pm  150^{k}$ & $20^{k}$ \\
N4594 & S0 & $29.79^{a}$ & $-24.87$ & $-22.24$ & 11.32 & 11.43 & 13.18 & $ 0.95\pm 0.09$ & $ 1900\pm  189^{j}$ & $40^{j}$ \\
M31 & Sb & $24.47^{m}$ & $-$ & $-21.80$ & 11.21 & 11.28 & 12.39 & $ 4.29\pm 0.00$ & $  450\pm  100^{n}$ & $30^{n}$ \\
N7331 & Sb & $30.59^{o}$ & $-24.68$ & $-21.84$ & 11.20 & 11.27 & 12.38 & $ 1.59\pm 0.37$ & $  210\pm  130^{p}$ & $49^{p}$ \\
N4552 & E & $30.92^{a}$ & $-24.25$ & $-21.29$ & 11.17 & 11.21 & 12.90 & $17.84\pm 0.62$ & $ 1400\pm  170^{e}$ & $50^{f}$ \\
N5128 & E & $27.90^{q}$ & $-24.13$ & $-21.48$ & 11.13 & 11.17 & 12.78 & $ 1.76\pm 0.00$ & $ 1550\pm  390^{r}$ & $48^{r}$ \\
N821 & E & $31.75^{a}$ & $-23.91$ & $-20.96$ & 11.04 & 11.06 & 12.52 & $ 0.95\pm 0.50$ & $  320\pm   45^{s}$ & $30^{s}$ \\
N3379 & E & $30.15^{t}$ & $-23.90$ & $-20.91$ & 11.03 & 11.05 & 12.51 & $ 4.12\pm 0.18$ & $  270\pm   68^{j}$ & $30^{j}$ \\
N1052 & E & $31.28^{a}$ & $-23.85$ & $-20.84$ & 11.01 & 11.03 & 12.49 & $ 1.80\pm 1.02$ & $  400\pm  120^{u}$ & $50^{u}$ \\
MW & Sbc & $- $ & $-$ & $-21.30$ & 10.99 & 11.02 & 12.46 & $ 4.19\pm 0.00$ & $  180\pm   20^{n}$ & $30^{n}$ \\
N1427 & E & $31.70^{a}$ & $-23.74$ & $-20.79$ & 10.97 & 10.99 & 12.43 & $ 4.94\pm 1.84$ & $  470\pm   80^{v}$ & $18^{v}$ \\
N1387 & S0 & $31.38^{a}$ & $-23.96$ & $-20.66$ & 10.95 & 10.97 & 12.41 & $ 5.80\pm 1.18$ & $  390\pm   27^{w}$ & $76^{w}$ \\
N1379 & E & $31.35^{a}$ & $-23.11$ & $-20.36$ & 10.72 & 10.73 & 12.09 & $ 5.79\pm 1.12$ & $  225\pm   23^{w}$ & $55^{w}$ \\
N7814 & Sab & $30.43^{a}$ & $-23.38$ & $-20.23$ & 10.69 & 10.70 & 12.04 & $ 0.91\pm 0.29$ & $  190\pm   20^{x}$ & $60^{x}$ \\
N3556 & Scd & $30.50^{y}$ & $-23.46$ & $-21.24$ & 10.67 & 10.71 & 12.02 & $ 1.77\pm 0.33$ & $  290\pm   80^{p}$ & $37^{p}$ \\
N4157 & Sb & $30.69^{y}$ & $-23.33$ & $-20.25$ & 10.66 & 10.67 & 12.00 & $ 7.55\pm 0.44$ & $   80\pm   20^{p}$ & $36^{p}$ \\
N2683 & Sb & $29.44^{a}$ & $-23.14$ & $-20.47$ & 10.58 & 10.59 & 11.88 & $ 2.48\pm 0.04$ & $  120\pm   40^{p}$ & $33^{p}$ \\
\hline
\end{tabular}
\end{table*}

\begin{table*}
 \centering
  \caption{Globular cluster specific frequencies.  See Table~\ref{variables} for parameter definitions.}\label{datatable2}
  \begin{tabular}{@{}lcccccccccc@{}}
\hline
Galaxy & T$_{N}$ &  T$_{blue}$ &  T$_{red}$ &  U$_{N}$ &  U$_{blue}$ &  U$_{red}$ &  V$_{N}$ &  V$_{blue}$ &  V$_{red}$ \\
\hline\hline
N4472 & $12.53\pm 1.53$ & $ 7.52\pm 0.92$ & $ 5.01\pm 0.61$ & $ 8.68\pm 1.06$ & $ 5.21\pm 0.64$ & $ 3.47\pm 0.42$ & $ 6.25\pm 0.76$ & $ 3.75\pm 0.46$ & $ 2.50\pm 0.31$ &  \\
N1407 & $15.97\pm 1.75$ & $ 9.58\pm 1.05$ & $ 6.39\pm 0.70$ & $12.17\pm 1.33$ & $ 7.30\pm 0.80$ & $ 4.87\pm 0.53$ & $ 8.76\pm 0.96$ & $ 5.26\pm 0.58$ & $ 3.50\pm 0.38$ &  \\
N4486 & $31.90\pm 2.13$ & $21.37\pm 1.42$ & $10.53\pm 0.70$ & $22.47\pm 1.50$ & $15.05\pm 1.00$ & $ 7.41\pm 0.49$ & $18.30\pm 1.22$ & $12.26\pm 0.82$ & $ 6.04\pm 0.40$ &  \\
N4374 & $ 5.22\pm 0.44$ & $ 3.65\pm 0.31$ & $ 1.56\pm 0.13$ & $ 3.98\pm 0.34$ & $ 2.79\pm 0.24$ & $ 1.20\pm 0.10$ & $ 3.25\pm 0.27$ & $ 2.28\pm 0.19$ & $ 0.98\pm 0.08$ &  \\
N1399 & $18.97\pm 2.29$ & $11.38\pm 1.37$ & $ 7.59\pm 0.92$ & $14.07\pm 1.70$ & $ 8.44\pm 1.02$ & $ 5.63\pm 0.68$ & $13.16\pm 1.59$ & $ 7.90\pm 0.95$ & $ 5.26\pm 0.64$ &  \\
N4406 & $10.26\pm 1.47$ & $ 6.16\pm 0.88$ & $ 4.11\pm 0.59$ & $ 8.05\pm 1.15$ & $ 4.83\pm 0.69$ & $ 3.22\pm 0.46$ & $ 7.84\pm 1.12$ & $ 4.70\pm 0.67$ & $ 3.13\pm 0.45$ &  \\
N3268 & $16.84\pm 0.53$ & $10.61\pm 0.34$ & $ 6.23\pm 0.20$ & $14.72\pm 0.46$ & $ 9.27\pm 0.29$ & $ 5.45\pm 0.17$ & $12.89\pm 0.41$ & $ 8.12\pm 0.26$ & $ 4.77\pm 0.15$ &  \\
N4636 & $16.42\pm 0.47$ & $10.57\pm 0.30$ & $ 5.85\pm 0.17$ & $14.28\pm 0.41$ & $ 9.19\pm 0.26$ & $ 5.08\pm 0.15$ & $14.56\pm 0.42$ & $ 9.37\pm 0.27$ & $ 5.18\pm 0.15$ &  \\
N3258 & $24.49\pm 0.61$ & $19.59\pm 0.49$ & $ 4.90\pm 0.12$ & $21.41\pm 0.54$ & $17.13\pm 0.43$ & $ 4.28\pm 0.11$ & $23.48\pm 0.59$ & $18.79\pm 0.47$ & $ 4.70\pm 0.12$ &  \\
N4594 & $ 9.14\pm 0.91$ & $ 5.48\pm 0.55$ & $ 3.66\pm 0.36$ & $ 7.01\pm 0.70$ & $ 4.21\pm 0.42$ & $ 2.80\pm 0.28$ & $12.53\pm 1.25$ & $ 7.52\pm 0.75$ & $ 5.01\pm 0.50$ &  \\
M31 & $ 2.77\pm 0.62$ & $ 1.94\pm 0.43$ & $ 0.83\pm 0.18$ & $ 2.38\pm 0.53$ & $ 1.67\pm 0.37$ & $ 0.71\pm 0.16$ & $18.20\pm 4.04$ & $12.74\pm 2.83$ & $ 5.46\pm 1.21$ &  \\
N7331 & $ 1.33\pm 0.82$ & $ 0.68\pm 0.42$ & $ 0.65\pm 0.40$ & $ 1.13\pm 0.70$ & $ 0.58\pm 0.36$ & $ 0.55\pm 0.34$ & $ 8.78\pm 5.43$ & $ 4.48\pm 2.77$ & $ 4.30\pm 2.66$ &  \\
N4552 & $ 9.42\pm 1.14$ & $ 4.71\pm 0.57$ & $ 4.71\pm 0.57$ & $ 8.70\pm 1.06$ & $ 4.35\pm 0.53$ & $ 4.35\pm 0.53$ & $17.73\pm 2.15$ & $ 8.86\pm 1.08$ & $ 8.86\pm 1.08$ &  \\
N5128 & $11.58\pm 2.91$ & $ 6.02\pm 1.52$ & $ 5.56\pm 1.40$ & $10.47\pm 2.63$ & $ 5.44\pm 1.37$ & $ 5.03\pm 1.26$ & $25.49\pm 6.41$ & $13.25\pm 3.33$ & $12.23\pm 3.08$ &  \\
N821 & $ 2.94\pm 0.41$ & $ 2.06\pm 0.29$ & $ 0.88\pm 0.12$ & $ 2.79\pm 0.39$ & $ 1.95\pm 0.27$ & $ 0.84\pm 0.12$ & $ 9.73\pm 1.37$ & $ 6.81\pm 0.96$ & $ 2.92\pm 0.41$ &  \\
N3379 & $ 2.51\pm 0.64$ & $ 1.76\pm 0.45$ & $ 0.75\pm 0.19$ & $ 2.40\pm 0.61$ & $ 1.68\pm 0.43$ & $ 0.72\pm 0.18$ & $ 8.35\pm 2.12$ & $ 5.84\pm 1.49$ & $ 2.50\pm 0.64$ &  \\
N1052 & $ 3.89\pm 1.17$ & $ 1.94\pm 0.58$ & $ 1.94\pm 0.58$ & $ 3.73\pm 1.12$ & $ 1.86\pm 0.56$ & $ 1.86\pm 0.56$ & $13.06\pm 3.92$ & $ 6.53\pm 1.96$ & $ 6.53\pm 1.96$ &  \\
MW & $ 1.84\pm 0.20$ & $ 1.29\pm 0.14$ & $ 0.55\pm 0.06$ & $ 1.70\pm 0.19$ & $ 1.19\pm 0.13$ & $ 0.51\pm 0.06$ & $ 6.26\pm 0.70$ & $ 4.38\pm 0.49$ & $ 1.88\pm 0.21$ &  \\
N1427 & $ 5.03\pm 0.86$ & $ 4.13\pm 0.70$ & $ 0.91\pm 0.15$ & $ 4.83\pm 0.82$ & $ 3.96\pm 0.67$ & $ 0.87\pm 0.15$ & $17.32\pm 2.95$ & $14.20\pm 2.42$ & $ 3.12\pm 0.53$ &  \\
N1387 & $ 4.36\pm 0.30$ & $ 1.05\pm 0.07$ & $ 3.31\pm 0.23$ & $ 4.21\pm 0.29$ & $ 1.01\pm 0.07$ & $ 3.20\pm 0.22$ & $15.18\pm 1.05$ & $ 3.64\pm 0.25$ & $11.53\pm 0.80$ &  \\
N1379 & $ 4.32\pm 0.44$ & $ 1.94\pm 0.20$ & $ 2.37\pm 0.24$ & $ 4.22\pm 0.43$ & $ 1.90\pm 0.19$ & $ 2.32\pm 0.24$ & $18.48\pm 1.89$ & $ 8.32\pm 0.85$ & $10.16\pm 1.04$ &  \\
N7814 & $ 3.89\pm 0.41$ & $ 1.56\pm 0.16$ & $ 2.33\pm 0.25$ & $ 3.82\pm 0.40$ & $ 1.53\pm 0.16$ & $ 2.29\pm 0.24$ & $17.14\pm 1.80$ & $ 6.85\pm 0.72$ & $10.28\pm 1.08$ &  \\
N3556 & $ 6.14\pm 1.69$ & $ 3.87\pm 1.07$ & $ 2.27\pm 0.63$ & $ 5.70\pm 1.57$ & $ 3.59\pm 0.99$ & $ 2.11\pm 0.58$ & $27.44\pm 7.57$ & $17.29\pm 4.77$ & $10.15\pm 2.80$ &  \\
N4157 & $ 1.75\pm 0.44$ & $ 1.12\pm 0.28$ & $ 0.63\pm 0.16$ & $ 1.72\pm 0.43$ & $ 1.10\pm 0.27$ & $ 0.62\pm 0.15$ & $ 7.94\pm 1.99$ & $ 5.08\pm 1.27$ & $ 2.86\pm 0.71$ &  \\
N2683 & $ 3.14\pm 1.05$ & $ 2.10\pm 0.70$ & $ 1.04\pm 0.35$ & $ 3.06\pm 1.02$ & $ 2.05\pm 0.68$ & $ 1.01\pm 0.34$ & $15.80\pm 5.27$ & $10.59\pm 3.53$ & $ 5.21\pm 1.74$ &  \\
\hline
\end{tabular}
\end{table*}

\begin{table}
 \centering
  \caption{Mass-to-light ratios used to compute galaxy masses. See text for derivation.}\label{tabml}
  \begin{tabular}{@{}lrrrrrrrrr@{}}
\hline
&E&S0&Sa&Sab&Sb&Sbc&Sc&Scd&Sd\\
\hline\hline
K-band &1.45&1.14&1.10&1.05&1.03&0.99&0.96&0.95&0.96\\
V-band &10.23&5.69&5.93&5.80&5.23&5.09&5.05&4.98&4.98\\
\hline
\end{tabular}
\end{table}

\label{lastpage}
\end{document}